\documentclass[11pt, a4paper]{article}

\usepackage{amsthm}
\usepackage{amsmath}
\usepackage{amssymb}
\usepackage{bbm}
\usepackage{booktabs}
\usepackage{natbib}
\usepackage{hyperref}
\usepackage{natbib}
\usepackage{graphicx}
\usepackage{color}
\usepackage[left = 2.5cm, right = 2.5cm, bottom = 3cm, top = 2.7cm]{geometry}
\usepackage{multirow}
\usepackage{subcaption}
\usepackage[export]{adjustbox}
\usepackage{calc}
\usepackage{bm}
\usepackage{pdfpages}
\usepackage{authblk}
\usepackage{wrapfig}

\newcommand*{\bb}{\boldsymbol}

\newcommand*{\SB}{S$\beta$}
\newcommand*{\VB}{V$\beta$}
\newcommand*{\MB}{M$\beta$}
\newcommand*{\MBA}{M$\beta$A}
\newcommand*{\FOMC}{FOMC}
\newcommand*{\MSTHP}{MSTHP}

\title{Flexible marked spatio-temporal point processes with applications 
to event sequences from association football}

\author[4]{Santhosh Narayanan}
\author[1]{Ioannis Kosmidis}
\author[2,3]{Petros Dellaportas}

\affil[1]{Department of Statistics, University of Warwick, Gibbet Hill Road, Coventry, CV4 7AL, UK}
\affil[2]{Department of Statistical Science, University College London, Gower St., London, WC1E 6BT, UK}
\affil[3]{Department of Statistics, Athens University of Economics and Business, 76 Patission Str., Athens, 10434, Greece }
\affil[4]{The Alan Turing Institute, 96 Euston Road, London, England, NW1 2DB, UK}

\begin{document}

\maketitle

\noindent

\begin{abstract}
  We develop a new family of marked point processes by focusing the
  characteristic properties of marked Hawkes processes exclusively to
  the space of marks, providing the freedom to specify a different
  model for the occurrence times. This is possible through the
  decomposition of the joint distribution of marks and times that
  allows to separately specify the conditional distribution of marks
  given the filtration of the process and the current time.  We
  develop a Bayesian framework for the inference and prediction from
  this family of marked point processes that can naturally accommodate
  process and point-specific covariate information to drive
  cross-excitations, offering wide flexibility and applicability in
  the modelling of real-world processes. The framework is used here
  for the modelling of in-game event sequences from association
  football, resulting not only in inferences about previously
  unquantified characteristics of game dynamics and extraction of
  event-specific team abilities, but also in predictions for the
  occurrence of events of interest, such as goals, corners or fouls in
  a specified interval of time.
\end{abstract}

\section{Introduction}
Football is one of the most popular team sports and is an example of
an invasive sport, where two opposing teams compete for the possession
of the ball with the dual objective of attacking to score a goal and
defending against attacks from the opposition. Most analyses in
football are typically done manually by studying video footage or
using simple frequency analysis of match events. Hence, there is huge
scope to improve the efficiency of the data-analytic methods as well
as the quality of performance evaluation. However, the analysis of
football data is mathematically and statistically challenging due to
the continuous interaction between players within and across the two
teams. As an introduction, we describe the event data from football
and survey the existing work in this area before arguing how marked
point processes are well-suited to developing a modelling foundation
to achieve our goal of describing the game dynamics.

\subsection{Football event data}
Over the last decade, the availability of spatio-temporal data from
team sports has inspired research into the application of statistical
methods for team and player performance evaluation. A comprehensive
survey of the recent research efforts into the spatio-temporal
analysis of team sports is provided in
\citet{gudmundsson2017spatio}. There are two primary types of
spatio-temporal data collected from team sports. Movement data
consists of samples of timestamped locations in the plane tracking the
movement of all players and the ball during the game. Player movement
is captured using fixed cameras in optical tracking systems, that
process the images to obtain the trajectories. Event data streams, on
the other hand, record the sequence of events that occur during the
game and are collected manually by trained analysts who watch video
feeds of the games through a special annotation software. As our work
is motivated by the availability of event data from football, we focus
on reviewing research that uses event data streams. Event data are
less dense than movement data, but richer in the sense that they
contain more information about what is happening in the game. Events
broadly fall into two categories; player events such as passes and
shots; and stoppage events such as fouls, end of game etc. Every event
is annotated with a timestamp, its location, its type (pass, foul,
etc.), the players involved, and team information.

A popular research topic based on event data is the network analysis
of player interaction. Models for player interaction can quantify a
team's playing style as well as the importance of an individual player
within the team. Players are identified as nodes of a network and are
connected using directed edges whose weights are proportional to the
number of successful passes between the two players. Passing networks
were first applied to team sports in \citet{passos2011networks} to
study a team's collective behaviour in water
polo. \citet{grund2012network} studied the degree centrality of
passing networks in football, which quantifies the importance of nodes
in the network based on the number of edges. They showed that teams
that rely heavily on key players performed relatively
worse. \citet{duch2010quantifying} used flow centrality to assess
player performance by capturing the fraction of times that a player
intervenes in those paths that result in a shot on goal. They also
take into account defensive behaviour by letting each player start a
number of paths proportional to the number of balls they
recover. \citet{clemente2015using} studied the density and
heterogeneity of passing networks and showed that high heterogeneity
leads to formation of sub-communities, meaning there is a low level of
cooperation between the players of a team. \citet{pena2012network}
looked at other centrality measures such as closeness and eigenvector
centrality as well as clustering in football passing networks.

Another use of event data is in the identification of frequently occurring 
sequences of passes between a small group of players within the same team. 
In \citet{borrie2002temporal}, passes are identified by the zones in the
pitch they start and end in and frequently occurring sequences are
detected by also taking into account the time intervals between
passes. \citet{wang2015discerning} proposed an unsupervised approach
to automatically detect tactical patterns in football. They present
the Team Tactic Topic model based on Latent Dirichlet Allocation to
identify tactics from pass sequences. Interesting visualisations are
provided for the most successful tactics as well as how a team's
tactical patterns evolve over a season. \citet{van2016strategy} also
look at automatic discovery of patterns in attacking strategy. They
use a data-driven approach to determine a number of spatial features
about the areas occupied during a continuous possession phase of a
team. The features are then used to cluster similar phases together to
identify frequently occurring event sequences within the
cluster. \citet{decroos2017predicting} partition the game using
overlapping intervals to create subsequences of events to use as a
feature to predict a goal event in the near future. They compute
similarity between subsequences using Dynamic Time Warping, a distance
measure for time-dependent sequences.

Extracting game states from event sequences to quantify the value of
player actions or to make predictions of the game outcome is another
interesting area of research. \citet{routley2015markov} used Markov
decision processes for valuing player actions in Ice Hockey. Game
states are derived from contextual features like game score and time
remaining along with the recent history of events. The associated
reward for an action in the Markov decision process gives the value of
the player action. A similar approach based on game states is taken in
\citet{decroos2018actions} to value player actions in football. They
train a classification model to calculate the probability a game state
will lead to a goal in the near future, where each game state is
described using over 150 features. The value of a player action is
then calculated by the shift in the predicted goal probability before
and after the action. Other approaches for predicting goal
probabilities based on a current game state are by
\citet{mackay2017predicting} and
\citet{robberechts2019will}. Approaches based on game states involve
significant effort into feature engineering, and with the use of
learning algorithms like gradient boosting that limit parameter
interpretations, the methods provide, typically, little insight into
the dynamics of the game.

The major focus of existing methods in team sport analysis appear to
be tailored towards individual player performance evaluation or
identifying specific patterns in team play. Most approaches take the
route of summarising the event data into compact representations like
networks and game states. In this paper, we take a more holistic
approach to study football as a dynamic system and model the entire
sequence of events within a game. Such a model, that captures all
event interactions, is attractive for predicting the occurrence of the
rare goal scored events, that determine the outcome of the game.

\subsection{Point processes}

Phenomena that are observed as a sequence of events happening over
time can be represented using point processes. While point processes
can describe a random collection of points in any general space, we
limit ourselves to the case in which the points denote events that
occur along a time axis. Such point processes, having a natural order
in which the points occur, are suitable for a wide range of real-world
applications and are well studied in probability theory.

As in \citet[Section~6.4]{daley2003theory}, processes in which points
are identified only by the occurrence times are referred to as
univariate point processes. Multivariate point processes, on the other
hand, are those in which the realisation of a discrete random
variable, say $m$, with a finite number of categories is recorded
along with the occurrence times. Marked point processes are processes
where $m$ is allowed to be a continuous random variable. An example
application of a marked point process with continuous marks is in
seismology, where the magnitude of an earthquake is recorded in
addition to the time of occurrence. In this paper, we model event
sequences observed in football using marked point processes with
discrete marks used to denote the event type.

When event sequence data are analysed using point process models, an
important distinction is between {empirical} models and
{mechanistic} models as noted by \citet{diggle2013spatial}. Empirical 
models have the solitary aim of describing the patterns in the observed 
data, while mechanistic models go beyond that and attempt to capture 
the underlying process that generated the data. Mechanistic models for 
marked point processes are typically specified using a joint conditional 
intensity for the occurrence times and the marks and in general are not 
flexible enough to be applied to complex real-world phenomena. The 
joint modelling of the components of the process can also be challenging 
and it is common to make strong restrictive assumptions like separability
\citep{gonzalez2016spatio} to simplify the model. In this paper, we
present a flexible mechanistic modelling framework for marked point
processes that are suitable for a wide range of applications without the 
need for assumptions like separability.

We produce a family of marked point processes that generalises the
classical Hawkes process, a mathematical model for {self-exciting}
processes proposed in \citet{hawkes1971spectra} that can be used to
model a sequence of arrivals of some type over time, for example,
earthquakes in \citet{ogata1998space}. Each arrival excites the
process in the sense that the chance of a subsequent arrival increases
for a period of time after the initial arrival and the excitation from
previous arrivals add up. Marked Hawkes processes are typically
specified using a joint conditional intensity function for the
occurrence times and the marks \citep[see, for example,][expression
2.2]{rasmussen2013bayesian}, and captures the magnitudes of all
cross-excitations between the various event types as well as the rate
at which these excitations decay over time. Excitation leads to
clustering of events in time as the process is driven by an intensity
that increases with every arrival for a short period of time. However,
in applications like the event sequences observed in football, the
events tend not to cluster in time and the marked Hawkes process model
is not suitable. The joint modelling of the times and the marks has to
be decoupled to restrict the excitation property of the process
exclusively to the dimension of the marks.

Similar to the decomposition of a multivariate distribution function
that motivated the partial likelihood in \citet{cox1975partial}, we
factorise the joint conditional distribution for a marked point
process into probability density functions for each event time
conditioned on the past occurrences times and marks, and probability
mass functions for the event marks conditioned on the time of
occurrence and the filtration of the process. Therefore, an
alternative approach to specify a marked point process model is to
specify the conditional distribution functions for the times and the
marks separately. We derive the conditional distribution function for
the marks from a marked Hawkes process, which gives us then the
freedom to specify the conditional distribution for the times
separately. In this way, we are able to construct marked point process
models that retain the characteristic properties of Hawkes processes,
such as excitation for the marks, while avoiding the strong clustering
of event times.

We develop a framework for Bayesian inference of such flexible marked
point processes, which is realised through the \textsf{Stan}
\citep{cmdstan} software for statistical modelling. \textsf{Stan}
implements a variant of the Hamiltonian Monte Carlo algorithm,
originally proposed by \citet{duane1987hybrid}, to generate samples
from the posterior distribution of the parameters. The Bayesian models
we consider are compared using the out-of-sample log predictive
density.

We define marked point processes for the modelling of touch-ball
events in football, which along with time and event type information
also carry location information. As we illustrate, the family of
marked point processes can be readily enriched to handle all times,
event types and locations. We are also able to incorporate team
information in a direct way that captures the relative abilities of
the teams for each event type. We develop a method based on
association rules \citep{agarwal1993mining} to reduce the complexity
introduced by the model extensions we introduce. The rule-based
approach identifies significant event interactions within sequences by
placing thresholds on particular measures of significance. We then
evaluate the accuracy of the excitation based models by comparing
against two baseline models and confirm the superior performance of
the models with excitation effects.

We provide a detailed parameter description showing how the model
parameters can be used to gain valuable insights into football. The
excitation component of the proposed model captures both the
magnitudes and the durations of all pairwise event interactions across
different locations. From the conversion rate parameters, we are able
to confirm the well-known home advantage effect, and quantify the
relative performance of each team when playing games at their home
venue compared to away. The conversion rate parameters are also driven
by team information, via the team ability parameters which, for
example, can capture the relative ability of a team to convert one
successful pass to another and retain possession of the ball. We also
discuss how the team ability parameters can be used to obtain rankings
for the teams by event type, that can be used as predictors of team
performance. The team ability parameters also capture some interesting
differences in the playing styles of the teams, that are not
immediately apparent just by looking at the event data. In this way,
the model along with its parameters can be used to develop a deeper
understanding of the game-play by coaching staff and inform strategic
decision making. The proposed model can also be used to simulate the
sequence of events in a game to obtain real-time predictions of event
probabilities. The simulator results in predictions that can enhance,
among others, the viewing experience of televised games. Finally, like
Hawkes Processes, the proposed model also allows the recovery of the
hidden branching structure of the process that quantifies the relative
contributions of the background process and previous occurrences to
the triggering of a new event.

The developments in this paper can be readily applied to many other
team sports like rugby, hockey, basketball etc. As none of the methods
have been tailored specifically to football or even sports for that
matter, they can also be applied to a wide range of applications that
generate event data streams.

\section{Data}

\subsection{Description and descriptives}
\label{sec:descriptives}

\begin{table}[t]
\centering
\caption{Events and their attributes from the first 20 seconds of the game 
between Southampton and West Ham United on September 15, 2013. For each 
event, we have records of the event time-stamp, team and player ids, event type, 
$(x,y)$ co-ordinates of its location in the playing field, and if the event type is a Pass, the event outcome (successful/unsuccessful) and the end $(x,y)$ co-ordinates.}
\resizebox{0.8\linewidth}{!}{
\begin{tabular}{rrrrllrrrr}
  \toprule
second & minute & team\_id & player\_id & type & x & y & outcome & end\_x & end\_y \\ 
  \midrule
 1 &  0 & 14 & 29544 & Pass & 50.1 & 48.8 & Successful & 51.1 & 48.2 \\ 
   2 &  0 & 14 & 21683 & Pass & 51.1 & 48.2 & Successful & 39.2 & 47.8 \\ 
   4 &  0 & 14 & 71714 & Pass & 39.2 & 47.8 & Successful & 29.5 & 77.6 \\ 
   6 &  0 & 14 & 118244 & Pass & 30.8 & 79.6 & Unsuccessful & 33.5 & 79.7 \\ 
  12 &  0 & 20 & 12533 & BallRecovery & 34.9 & 89.9 &  &  &  \\ 
  13 &  0 & 20 & 12533 & Pass & 35.9 & 88.3 & Successful & 37.3 & 76.1 \\ 
  15 &  0 & 20 & 8247 & Pass & 34.9 & 77.0 & Unsuccessful & 44.9 & 85.9 \\ 
  16 &  0 & 14 & 71714 & Interception & 53.2 & 16.7 &  &  &  \\ 
  18 &  0 & 14 & 69375 & Pass & 43.1 & 23.1 & Unsuccessful & 70.9 & 9.7 \\ 
  \bottomrule
\end{tabular}}
\label{tab:datasnapshot}
\end{table}

The data that motivated this work was provided by Stratagem
Technologies Ltd, and consists of all touch-ball events from all
English Premier League games in the 2013/14 season. A touch-ball event
is an event where a player has acted on the ball by touching it with
some part of their body. We identified mistakes in the original data,
with the most critical issues relating to impossible sequences of
consecutive events (e.g. a dribble a few seconds after a goal). Such
data issues have been addressed in a systematic way, using the
data-cleaning workflow in the publicly-available PhD thesis by
Narayanan \citep[see,][Section~5.3]{narayanan:2021}. In total, the
data consists of over half a million touch-ball events recorded over
the season along with other attributes.  A snapshot of the data is
provided in Table \ref{tab:datasnapshot}. The league is contested by a
total of 20 teams and follows a round-robin tournament scheduling,
where each team plays every other team at their home and away venues,
which results in a total of 380 games over the season.

\begin{table}[t]
\centering
\caption{Frequencies of the 22 distinct types of touch-ball events in the data.}
\resizebox{0.6\linewidth}{!}{
\begin{tabular}{lrlr}
  \toprule
event type & frequency & event type & frequency \\ 
  \midrule
Pass & 376924 & SavedShot & 4971 \\ 
  BallRecovery & 36908 & Save & 4910 \\ 
  Clearance & 25462 & CornerAwarded & 4100 \\ 
  Tackle & 14581 & MissedShots & 4076 \\ 
  TakeOn & 13607 & OffsidePass & 1582 \\ 
  BallTouch & 13517 & Claim & 1181 \\ 
  Aerial & 12871 & Goal & 1052 \\ 
  Interception & 10422 & Punch & 380 \\ 
  Dispossessed & 8897 & ShotOnPost & 187 \\ 
  Foul & 8238 & Smother & 122 \\ 
  KeeperPickup & 5208 & CrossNotClaimed & 81 \\ 
   \bottomrule
\end{tabular}}
\label{tab:eventfreq}
\end{table}

Each game comprises two halves that are separated by an interruption
of approximately 15 minutes. In what follows, we refer to each
uninterrupted game half as a game period. For each touch-ball event,
we have records of the event type, time-stamp, $(x,y)$ co-ordinates of
its location in the playing field, team and unique player identifiers,
game period, and if the touch-ball event is a Pass, the event outcome 
(successful/unsuccessful) and the end $(x,y)$ co-ordinates . Table
\ref{tab:eventfreq} gives the frequency of each of the 22 distinct
touch-ball event types recorded in the data.

\begin{figure}[!t]
\centering
\resizebox{0.8\linewidth}{!}{
\includegraphics{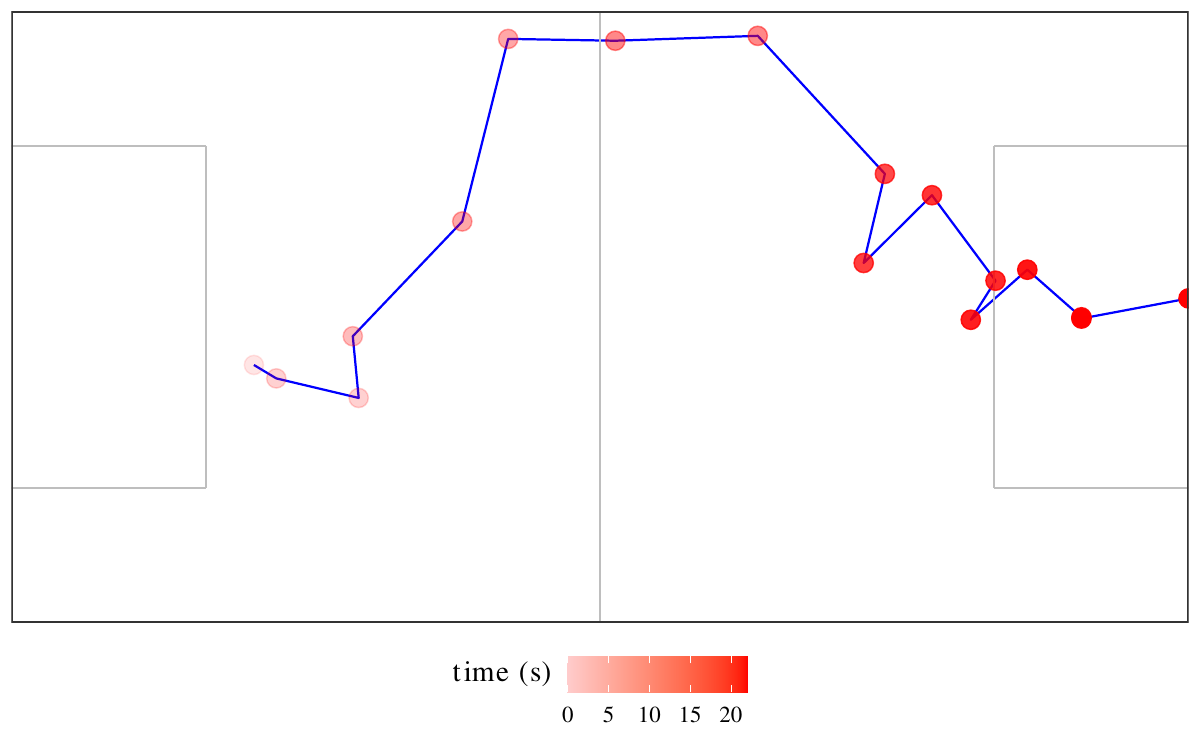}}
\caption{Tracing the locations of the sequence of events that led to the 
goal scored by Jack Wilshere for Arsenal against Norwich City (voted the 
best goal of the 2013/14 season).}
\label{fig:balltrack}
\end{figure}

Figure~\ref{fig:balltrack} shows the trajectory of the ball during an
attacking move that led to a goal in the 18th minute of the game
between Arsenal and Norwich City on October 19, 2013. The goal was
scored by Jack Wilshere for Arsenal and was voted as the best goal 
in the English Premier League for the 2013/14 season.

\begin{figure}[!t]
\centering
\resizebox{0.8\linewidth}{!}{
\includegraphics{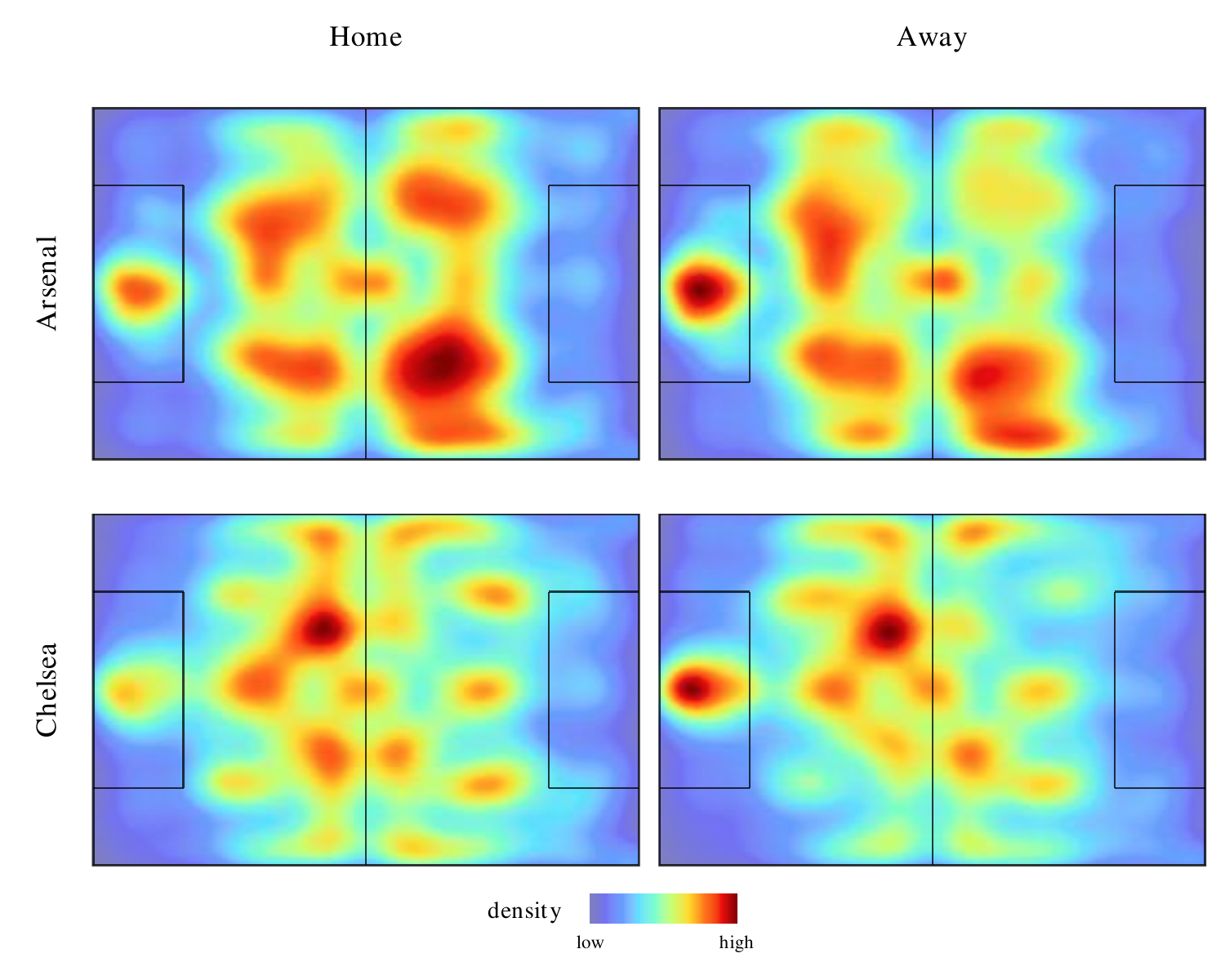}}
\caption{Heat maps showing the density of ball-touches for Arsenal and
  Chelsea in their home and away games in the 2013/14 season. In all
  heat maps the team is attacking to the right.}
\label{fig:homeaway}
\end{figure}

Latent game characteristics, such as the home advantage, and
differences in playing styles and formations between the teams are
also reflected in the touch-ball events. For example,
Figure~\ref{fig:homeaway} compares the concentration of ball touches
for Arsenal and Chelsea between their home and away games in the
2013/14 season. The playing field is plotted so that the team is
always attacking to the right. It is clear that when the teams play at
home the density of events is higher towards the opponent's goal. 
In fact, the point process modelling framework developed in this paper
allows us to quantify home advantage by, for example, learning the 
relative ability of each team to retain possession when playing at 
home compared to away (see Section~\ref{homeadvref}).

\subsection{Data preparation}

\begin{table}[t]
\centering
\caption{Composite event types along with their labels and observed frequencies in the data.}
\resizebox{0.6\linewidth}{!}{
\begin{tabular}{llr}
  \toprule
m & mark label & count \\ 
  \midrule
  1 & Home\_Win & 10864 \\ 
    2 & Home\_Dribble & 3432 \\ 
    3 & Home\_Pass\_S & 152140 \\ 
    4 & Home\_Pass\_U & 42344 \\ 
    5 & Home\_Shot & 5127 \\ 
    6 & Home\_Keeper & 3273 \\ 
    7 & Home\_Save & 2208 \\ 
    8 & Home\_Clear & 11780 \\ 
    9 & Home\_Lose & 16534 \\ 
   10 & Home\_Goal & 597 \\ 
   11 & Home\_Foul & 4229 \\ 
   12 & Home\_Out\_Throw & 8982 \\ 
   13 & Home\_Out\_GK & 3084 \\ 
   14 & Home\_Out\_Corner & 2321 \\ 
   15 & Home\_Pass\_O & 814 \\ 
  \bottomrule
\end{tabular}
\begin{tabular}{llr}
  \toprule
m & mark label & count \\ 
  \midrule
   16 & Away\_Win & 10829 \\ 
   17 & Away\_Dribble & 3123 \\ 
   18 & Away\_Pass\_S & 140975 \\ 
   19 & Away\_Pass\_U & 41462 \\ 
   20 & Away\_Shot & 4107 \\ 
   21 & Away\_Keeper & 3555 \\ 
   22 & Away\_Save & 2702 \\ 
   23 & Away\_Clear & 14059 \\ 
   24 & Away\_Lose & 16515 \\ 
   25 & Away\_Goal & 455 \\ 
   26 & Away\_Foul & 4009 \\ 
   27 & Away\_Out\_Throw & 8396 \\ 
   28 & Away\_Out\_GK & 3697 \\ 
   29 & Away\_Out\_Corner & 1779 \\ 
   30 & Away\_Pass\_O & 768 \\ 
   \bottomrule
\end{tabular}}
\label{tab:encmarks}
\end{table}

We combine the types and outcomes of touch-ball events into
in-play and terminal composite events. The in-play composite events
are Win, Dribble, Successful Pass (Pass\_S), Unsuccessful Pass
(Pass\_U), Shot, Keeper, Save, Clear and Lose. Win denotes a player
regaining possession of the ball from the opponent. Dribble is taking
the ball forward with repeated slight touches. Passes are deemed to 
be successful when the ball is received by a teammate and 
unsuccessful otherwise. Shots include all attempts on the opponents' 
goal, including those missing the target. The Keeper event denotes 
the goal keeper taking possession of the ball into their hands by picking 
it up or claiming a cross. The Keeper event is unlike any other in-play 
event, as the goal keeper is allowed to hold the ball without being 
challenged for a period of time while waiting for opponents to clear the 
goal area. As a result, there is often a delay before the next event even 
though the ball is technically in-play. Saves are events where the goal 
keeper prevents a shot from crossing the goal line. Clear events are 
those where a player moves the ball away from their goal area to 
safety while the Lose event is when a player loses possession of the ball. 
The terminal composite events are those which result in the ball going 
out-of-play and are Goal, Foul, Out\_Throw, Out\_GK, Out\_Corner and 
Offside Pass (Pass\_O). The terminal events interrupt the game resulting 
in a delay before play resumes. Each composite event is tracked for both 
the home and away teams. For this reason, we append ``Home'' or 
``Away'' as a prefix to the event label to distinguish between the events 
of the two teams playing the game. This results in $M = 30$ distinct 
composite events, whose labels and observed frequencies are given in 
Table~\ref{tab:encmarks}.

\begin{figure}[!t]
\centering
\resizebox{0.6\linewidth}{!}{
\includegraphics{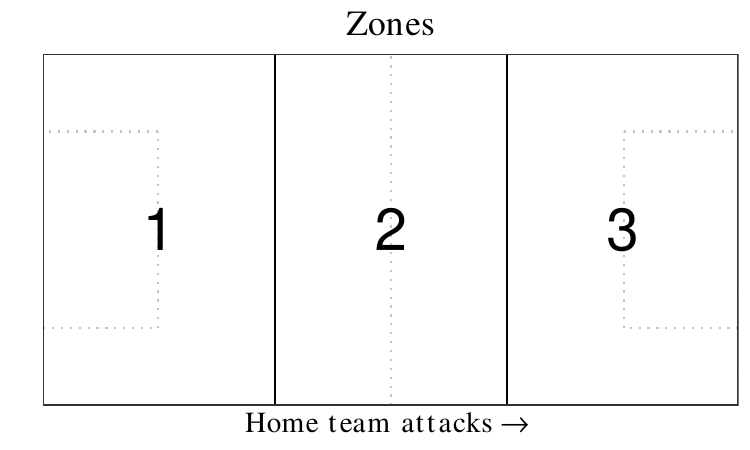}}
\caption{Mapping from event location in $(x,y)$ coordinates to zones.}
\label{fig:zones}
\end{figure}

For each touch-ball event, the data also contains the associated $(x,y)$
coordinates on the playing field. We partition the playing field into 3
zones of equal area. The zones and their corresponding labels are shown
in Figure~\ref{fig:zones}. Zone 1 is the region where the home team
defends their goal, zone 3 is the region where the home team attacks,
and zone 2 is the midfield region. It is natural to expect that the control
a team has on the game depends on the zone the ball is at. For
example, the home team is expected to retain possession of the ball
more successfully in zone 1 as compared to zone 3. 

Table~\ref{tab:dtsnap} shows a snapshot of the data after its preparation, 
including a unique identifier for each game period in the data.

\

\begin{table}[t]
\centering
\caption{Snapshot of the final data prepared for modelling where each event, indexed by $i = 1,...,n$, consists of the following components, the time of occurrence $t_i$, the zone $z_i$, and the mark $m_i$. The home and away team information for each game is assumed to be known and the first event ($t_1, z_1, m_1$) in each game period is considered to be deterministic and therefore, not modelled.}
\resizebox{0.6\linewidth}{!}{
\begin{tabular}{rrrrrrrrr}
  \toprule
i & id & period & team\_id & time ($t_i$) & zone ($z_i$) & mark ($m_i$)\\ 
  \midrule
1 &  101 & 1 & 1 & 0 & 2 & 18 \\ 
2 &  101 & 1 & 1 & 1 & 2 & 19 \\ 
3 &  101 & 1 & 2 & 3 & 1 & 8 \\ 
4 &  101 & 1 & 1 & 6 & 3 & 16  \\ 
5 &  101 & 1 & 1 & 8 & 3 & 18  \\ 
6 &  101 & 1 & 1 & 15 & 2 & 18  \\ 
7 &  101 & 1 & 1 & 16 & 1 & 19 \\ 
8 &  101 & 1 & 2 & 19 & 1 & 12 \\ 
   \bottomrule
\end{tabular}}
\label{tab:dtsnap}
\end{table}

\section{Marked point processes}

\subsection{Conditional intensity function}

Sequences of events over time are conveniently represented as
realisations of a point process. Oftentimes, the events can carry
additional information, which are assumed to be realisations of random
variables, referred to as marks. The collection of the times $\{t_i\}$
at which the events occur and the marks $\{m_i\}$ is a marked point
process, whose ground process, is the process for $\{t_i\}$ only.

A marked point process is typically specified through its joint conditional 
intensity function
\begin{equation}
\label{exp:mppint}
\lambda^*(t, m) = \lambda^*_g(t)f^*(m \mid t) \, ,
\end{equation}
where $\lambda^*_g(t)$ is the conditional intensity of the ground
process and $f^*(m \mid t)$ is the conditional probability density or
mass function of the mark $m$ at time $t$. Both $\lambda^*_g(t)$ and
$f^*(m \mid t)$ in~(\ref{exp:mppint}) are understood as being
conditional on $\mathcal{F}_{t^-}$, which is the filtration of the
marked point process up to but not including $t$.

\subsection{Marked Hawkes processes}
\label{sec:markhpdef}

Marked Hawkes processes are point processes whose defining
characteristic is that they {self-excite}, meaning that each arrival
increases the rate of future arrivals for a period of time. More
formally, consider a realisation of a marked point process, consisting
of event times $\{t_i\}$ with $t_i \in \mathbb{R}^+$ and
$t_i > t_{i-1}$, and marks $m_i \in \{1, \ldots, M\}$
$(i = 1, \dots, n)$, where $M$ is the number of discrete marks. The
marked Hawkes process is most intuitively specified using its mark
dependent conditional intensity function $\lambda^*(t, m)$, which for
an exponentially decaying intensity is \citep[expression
2.2]{rasmussen2013bayesian},
\begin{equation}
\lambda^*(t, m) = \mu\delta_{m} + \sum_{t_j < t} \epsilon\beta \mathrm{e}^{-\beta (t-t_j)}\gamma_{m_{j} \to m} \, .
\label{exp:hawkint}
\end{equation}
In~(\ref{exp:hawkint}), the parameter $\mu > 0$ is a constant
background intensity and $\delta_{m} \in (0,1)$ is the background mark
probability for mark $m$ with $\sum_{m = 1}^M \delta_m = 1$. The
parameter $\epsilon \in (0,1)$ is the excitation factor, $\beta > 0$ is
the exponential decay rate and $\gamma_{m_{j} \to m} \in (0,1)$ is the
probability the excitation from an event of mark $m_j$ triggers an
event of mark $m$, with $\sum_{m = 1}^M \gamma_{m_{j} \to m} = 1$ for
any $m_j \in \{1, \ldots, M\}$.

\subsection{Limitations of the marked Hawkes process model}
\label{sec:lims}

\begin{figure}[!t]
\centering
\resizebox{\linewidth}{!}{
\includegraphics{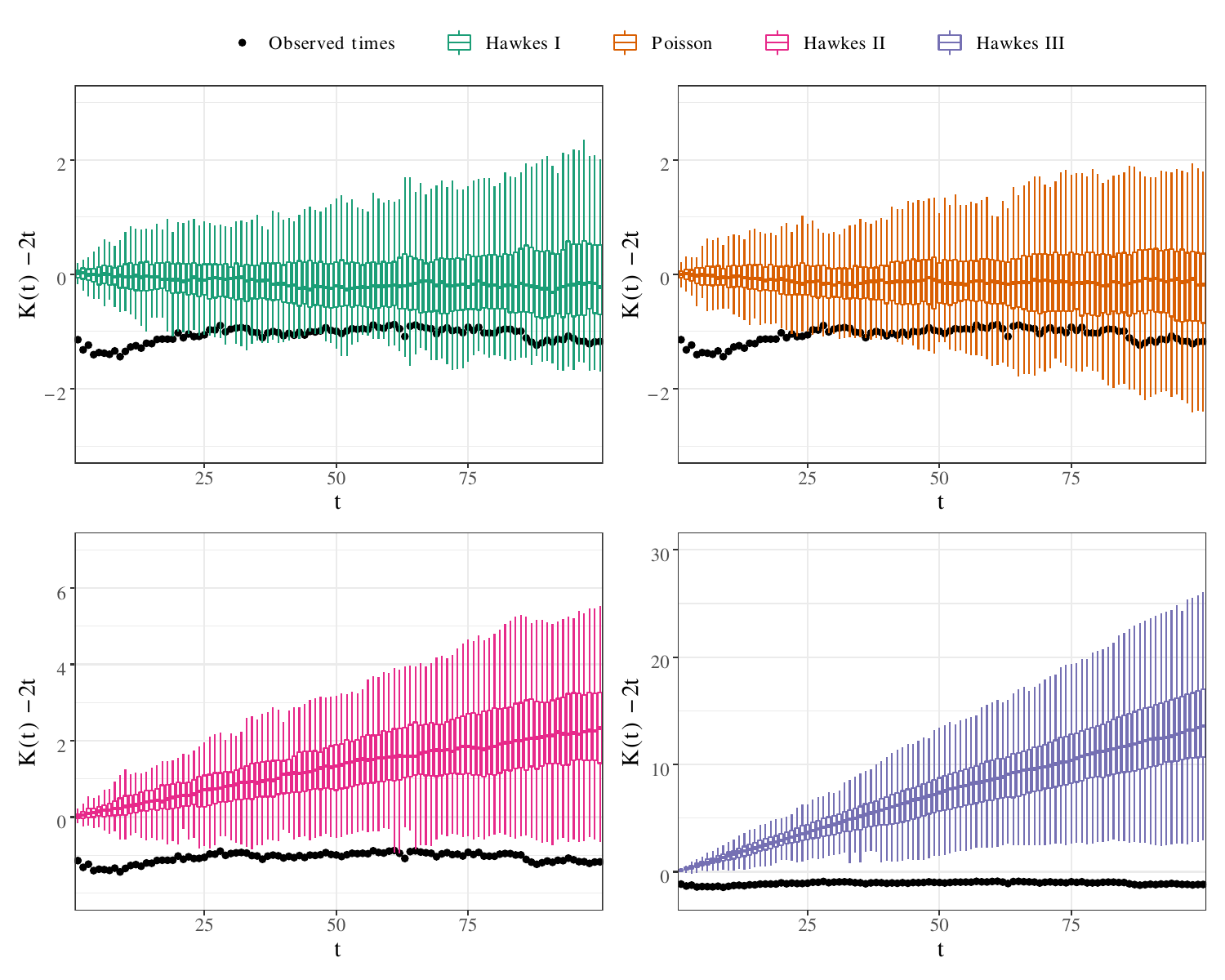}}
\caption{The K function estimate $\hat{K}(t) - 2t$ of the observed event times (black points) from the first game of the season between Aston Villa and Arsenal. Hawkes I (green) is a Hawkes process with parameters $(\mu, \epsilon, \beta) = (0.4183, 0.0035, 0.0004)$ estimated from the observed times using maximum likelihood. A Hawkes process with $\epsilon = 0$ is the trivial case with no excitation that reduces to a Poisson process (orange) with estimated rate $\mu = 0.4189$. Hawkes II (pink) has parameters $(\mu, \epsilon, \beta) = (0.2594, 0.4, 0.01)$ and Hawkes III (purple) has parameters $(\mu, \epsilon, \beta) = (0.1068, 0.8, 0.01)$. Hawkes II and Hawkes III are examples of processes with moderate and severe clustering respectively, whose $\mu$ parameters were estimated from the observed times using maximum likelihood after fixing $\epsilon, \beta$. The box plots of the estimates for the Hawkes and Poisson processes were computed using $100$ independent simulations of each process over the same time interval as the observed times.}
\label{fig:kfuncs}
\end{figure}

\begin{figure}[!t]
\centering
\resizebox{0.7\linewidth}{!}{
\includegraphics{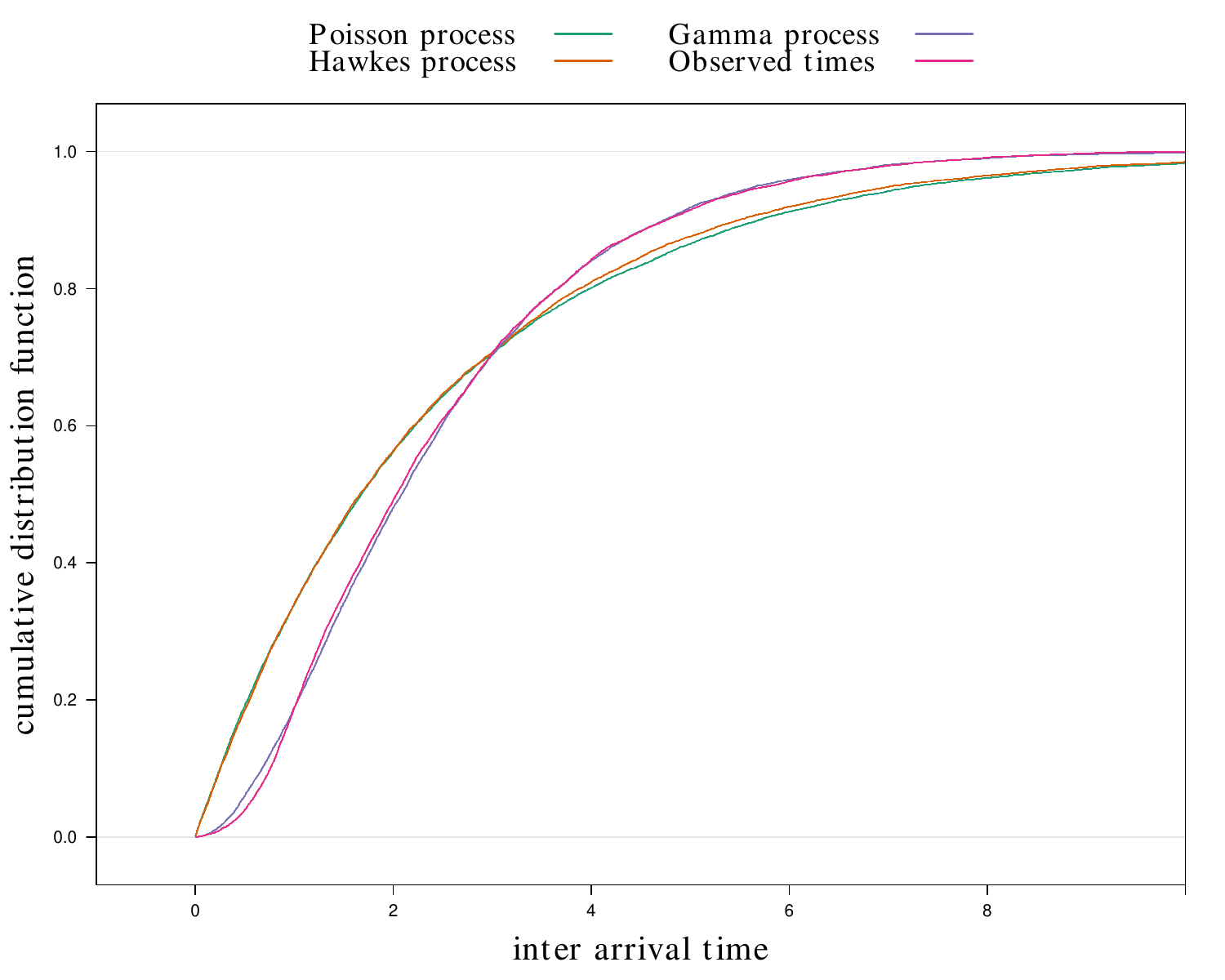}}
\caption{Comparing the cumulative distribution functions (CDFs) of the inter-arrival times of events simulated from a Poisson process (green), a Hawkes process (orange), a Gamma process (purple) and observed event times in football (pink). Empirical CDFs were computed using $10,000$ inter-arrival times in each case.}
\label{fig:cdfcluster}
\end{figure}

The specification in~(\ref{exp:hawkint}) describes a marked Hawkes
process that is linear in the sense that the excitations from
different arrivals add up, not only increasing the probability of triggering 
an event of a particular type, but also concentrating the occurrence 
times for a certain period of time. For this reason, marked Hawkes 
processes have proven useful in a wide range of applications, where 
events tend to cluster in time, such as the modelling of earthquakes 
\citep{ogata1998space}, gang violence \citep{mohler2011self} and 
financial market events \citep{bowsher2007modelling}.

However, in applications like the modelling of event sequences in
football, each event triggers other events of a particular type
with high probability, while it is not necessarily true that the occurrence
times cluster.

As an illustration that the observed events in football do not exhibit
clustering, consider only the collection of the times $\{t_i\}$ at
which the events occur. A succinct method to investigate the
aggregation of the points is using the non-parametric Ripley's
K-function summary \citep{ripley1977modelling}, which is the reduced
second moment measure. An estimator of the K-function in the
one-dimensional case is derived in \citet{diggle1985kernel} as
\begin{equation}
\hat{K}(t) = \frac{T}{n^2} \sum_{i=1}^{n}\sum_{j \neq i}
w_{ij} \mathbbm{1}(|t_i - t_j| \leq t) \, ,
\end{equation}
where $(0,T)$ is the time interval over which the $n$ points are
observed, $\mathbbm{1}(.)$ is the indicator function, and $w_{ij}$ is
an edge correction taking values $w_{ij} = 1$ if
$|t_i - t_j| \leq \min(t_i, T - t_i)$ and $w_{ij} = 2$, otherwise.

For a homogeneous Poisson process, $K(t) = 2t$. If $K(t) > 2t$, the
process is said to be \textit{over-dispersed} relative to the Poisson,
and exhibits some degree of clustering. If $K(t) < 2t$, the process is
\textit{under-dispersed} relative to the Poisson, and tends towards
regular occurrences. Figure~\ref{fig:kfuncs} shows the K function
estimate, $\hat{K}(t) - 2t$ for $t \in \{1, 2, \ldots, 100\}$, of the
observed event times from the first game of the season between Aston
Villa and Arsenal ($n = 1279$).  We also compare the estimates from
those observed times with the estimates of the events simulated from
several one-dimensional Hawkes processes with conditional intensity of
the form,
$\lambda^*(t) = \mu + \sum_{t_j < t} \epsilon\beta \mathrm{e}^{-\beta
  (t-t_j)}$.  Hawkes I (green) is the fitted Hawkes process with
parameters $(\mu, \epsilon, \beta) = (0.4183, 0.0035, 0.0004)$
estimated from the observed times using maximum likelihood. Note that
the estimated $\epsilon$ is very close to 0, indicating no
clustering. A Hawkes process with $\epsilon = 0$ is the trivial case
with no excitation that reduces to a Poisson process (orange) with
estimated rate $\mu = 0.4189$. Hawkes II (pink) has parameters
$(\mu, \epsilon, \beta) = (0.2594, 0.4, 0.01)$ and Hawkes III (purple)
has parameters $(\mu, \epsilon, \beta) = (0.1068, 0.8, 0.01)$. Hawkes
II and Hawkes III are examples of processes with moderate and severe
clustering respectively, whose $\mu$ parameters were estimated from
the observed times using maximum likelihood after fixing
$\epsilon, \beta$. The box plots of the estimates for the Hawkes and
Poisson processes were computed using $100$ independent simulations of
each process over the same time interval as the observed times.

The $\hat{K}(t) - 2t$ values for the Hawkes II and Hawkes III
processes quickly get above 0 and increase with $t$ demonstrating the
behaviour of processes with different degrees of clustering. On the
other hand, the $\hat{K}(t) - 2t$ values for the fitted Hawkes process
(Hawkes I) and the Poisson process concentrate around 0, being
indicative of the expected behaviour of processes where points do not
cluster. The $\hat{K}(t) - 2t$ values from the observed times range
from -1.4 to -0.9 indicating slight under-dispersion relative to the
Poisson process. In other words, the observed times in football
exhibit no clustering and in fact show evidence of being more regular
than the Poisson process.

Another method to investigate the aggregation of points is by looking at 
distribution of the inter-arrival times. Figure~\ref{fig:cdfcluster} shows the 
empirical distribution function of the first $10,000$ inter-arrival times from 
the first 7 games of the  league season. We also plot the cumulative 
distribution functions of a homogeneous Poisson process and a  
Hawkes process, whose parameters are estimated from the aforementioned 
$10,000$ events using maximum likelihood. The fitted Hawkes process is far 
from the empirical distribution function, and almost identical to the fitted
Poisson process confirming that the arrival times in football do not cluster. 
This is further evidence that Hawkes processes are not appropriate for 
modelling events such as those observed in football, that tend not to 
cluster in time. Figure~\ref{fig:cdfcluster} also includes the cumulative 
distribution function of a fitted Gamma process, which, as is apparent 
provides an excellent fit to the observed inter-arrival times.

\section{Specification of flexible marked point processes}
\label{sec:spec}

\subsection{Decoupling the modelling of times and marks}
\label{sec:frameworkref}

According to the decomposition of a multivariate distribution function
in \citet[expression 2]{cox1975partial} the likelihood of a marked
point process observed in $(0, T)$ can always be factorised as
\begin{equation}
\mathcal{L}(\mathcal{F}_{t_{n}} \mid \bb{\zeta}, \bb{\theta}) = \prod_{i = 1}^n \left\{ g(t_i \mid \mathcal{F}_{t_{i-1}}; \bb{\zeta})  f(m_i \mid t_i, \mathcal{F}_{t_{i-1}}; \bb{\theta}) \right\} \left\{1 - G(T \mid \mathcal{F}_{t_{n}}; {\bb\zeta})\right\}  \, ,
\label{exp:cox}
\end{equation}
where $g$, $G$, and $f$ are the conditional density and distribution
function for the times, and the probability mass function for the
marks, respectively, and $\bb{\zeta}, \bb{\theta}$ are unknown
parameter vectors that may or may not share components. The last term
in~(\ref{exp:cox}) accounts for the fact that the unobserved occurrence
time $t_{n+1}$ must be after the end of the observation interval
$(0, T)$. Therefore, an alternative approach to specify a marked point
process is to specify the functions
$g(\cdot \mid \mathcal{F}_{t_{i-1}}; \bb{\zeta})$ and
$f(\cdot \mid t_i, \mathcal{F}_{t_{i-1}}; \bb{\theta})$, separately,
and combine them as in~(\ref{exp:cox}). The key insight in the current
work is to derive the specification for the marks
$f(\cdot \mid t_i, \mathcal{F}_{t_{i-1}}; \bb{\theta})$ from the joint
conditional intensity function of a marked Hawkes process model, and
then to specify a probability density function for the times
$g(\cdot \mid \mathcal{F}_{t_{i-1}}; \bb{\zeta})$ best suited to our
application. In this way, we can restrict the characteristic
excitation property of marked Hawkes processes exclusively to the
modelling of the marks, and have the freedom to specify a different
model for the occurrence times.

By the definition of the conditional intensity function for a marked
point process in~(\ref{exp:mppint}),
$f(m_i \mid t_i, \mathcal{F}_{t_{i-1}}; \bb{\theta}) = \lambda^*(t_i,
m_i) / \sum_{m = 1}^M \lambda^*(t_i, m)$. Plugging in $\lambda^*(t_i, m)$
from~(\ref{exp:hawkint}) in the latter expression, gives
\begin{equation}
f(m_i \mid t_i, \mathcal{F}_{t_{i-1}}; \bb{\theta}) = \frac{\delta_{m_i} + \sum_{t_j < t_i} \alpha^* \mathrm{e}^{- \beta (t_i-t_j)}\gamma_{m_{j} \to m_{i}}}{1 + \sum_{t_j < t_i} \alpha^* \mathrm{e}^{- \beta (t_i-t_j)}} \, ,
\label{exp:modspecv1}
\end{equation}
where $\alpha^* = \frac{\epsilon\beta}{\mu}$. 
Expression~(\ref{exp:modspecv1}) makes it immediately apparent, that the 
parameters  $\mu$ and $\epsilon$ of the marked Hawkes process as specified 
by~(\ref{exp:hawkint}) are not always identifiable for general specifications 
of $g(\cdot \mid \mathcal{F}_{t_{i-1}}; \bb{\zeta})$ in~(\ref{exp:cox}). Apart 
from a mathematical fact, this is also rather intuitive, because $\mu$ and 
$\epsilon$ in~~(\ref{exp:hawkint}) characterise the evolution of the Hawkes 
process in the time dimension and the sequence of marks is not sufficient 
to identify them.

The specification of the marked point process likelihood is complete once a 
probability density function $g(\cdot \mid \mathcal{F}_{t_{i-1}}; \bb{\zeta})$ 
for the event times is specified.

We should highlight here that the marked point processes from the 
factorisation in~(\ref{exp:cox}) are generally different to the ones that 
result by assuming separability of the conditional intensity functions 
\citep[see, for example,][Section 6.5]{gonzalez2016spatio}. A separable 
conditional intensity functions has the form
\begin{equation*}
\lambda^*(t, m) = \lambda^*_g(t) f^*(m) \, .
\end{equation*}
and implies that the conditional distribution of the mark does not depend 
on the occurrence time $t$. Separability is a convenient assumption because 
it allows for the sequence of marks to be modelled separately from the 
sequence of times. In contrast, the factorisation in~(\ref{exp:cox}) allows 
the conditional distribution of the mark to depend on the time of occurrence 
as well as the history, still allowing for estimating ${\bb \theta}$ separately 
from ${\bb \zeta}$, if ${\bb \theta}$ and ${\bb \zeta}$ do not share 
components.

The proposed marked point process model also allows the recovery of
the hidden branching structure of the process, a key feature of Hawkes
Processes \citep{hawkes1974cluster}. In Section~\ref{sec::eventgene},
we calculate the branching structure probabilities and quantify the
relative contributions of the background process and previous
occurrences to the triggering of a new event.

\subsection{Parameter interpretation}
\label{sec:modparinterpret}

In~(\ref{exp:modspecv1}), the mark probability of each event in the 
sequence is determined by a combined additive effect from a background 
component and all previous occurrences. The first term $\delta_{m_i}$ in 
the numerator is the mark probability associated with the background 
component, while each term 
$\alpha^* \mathrm{e}^{- \beta (t_i-t_j)}\gamma_{m_{j} \to m_{i}}$ is the 
contribution from the excitation caused by a previous occurrence in the 
sequence.

The background mark probability $\delta_{m} \in (0,1)$ is the probability 
an event has a mark $m$ if the event is triggered solely by the background 
component. The excitation factor $\alpha^* \geq 0$ is a scaling factor 
applied to the contributions from the previous occurrences to the event 
mark probability. Large values of $\alpha^*$ indicate a stronger 
dependence of the process on its history, because the contributions from 
previous occurrences are weighted higher relative to the background 
component. The decay rate $\beta > 0$ is the exponential rate at which 
the excitations from previous occurrences decay over time. The parameter 
$\gamma_{m_{j} \to m_i} \in (0, 1)$ is the probability the excitation from 
an event of mark $m_j$ triggers an event of mark $m_i$. In other words, 
$\gamma_{m_{j} \to m_i}$ can be viewed as the conversion rate for the 
transition from an event with mark $m_{j}$ to an event with mark $m_i$.

In summary, as in marked Hawkes processes, the specification for the marks 
in~(\ref{exp:modspecv1}) captures not only all cross-excitations between 
the various marks but also the rate at which these excitations decay over 
time.

\subsection{Covariate-driven cross excitation}
\label{sec:covariates}

The conditional distribution of marks with probability mass 
function~(\ref{exp:modspecv1}), allows to drive the cross-excitation of 
the marks using covariates. The conversion rates $\gamma_{m_{j} \to m}$ 
can be linked to a covariate vector 
${\bb x} = (x_{1}, \ldots, x_{p})^\top$ observed at the current time 
through the baseline-category logit specification 
\citep[see, for example,][Section 6.1]{agresti2007categorical} 
\begin{equation}
\log\left(\frac{\gamma_{m_{j} \to m}}{\gamma_{m_{j} \to M}}\right) =  \phi_{m_{j} \to m} + {\bb \omega}_{m}^\top {\bb x} \qquad (m = 1,\ldots, M-1) \, ,
\label{exp:baselogitspecbayes}
\end{equation}
where ${\bb \omega}_m$ is an unknown $p$-vector of regression
parameters. Then, keeping all covariates apart from $x_{t}$ fixed,
$\omega_{mt}$ is the log of the ratio of odds for category $m$ versus
the baseline category $M$ at $x_{t} + 1$ to that at $x_{t}$
$(t = 1, \ldots, p)$. Also, by setting all covariates $x_{t}$ equal to
0, $\phi_{m_{j} \to m_i}$ is the log of the ratio of odds for category
$m$ versus the baseline category $M$. The covariate vector ${\bb x}$
can include a combination of process-specific covariates that are
time-invariant, and event-specific covariates.  For example, in
Section~\ref{sec:fittedmodels}, we use~(\ref{exp:baselogitspecbayes})
to parameterise the marked point process in terms of the relative
abilities of teams for each event type, and produce team rankings per
event-type.

\subsection{Spatio-temporal marked point processes}

We can readily extend the factorisation of the likelihood
in~(\ref{exp:cox}) to include conditional densities for the event
locations, when the latter are observed. We can write
\begin{equation}
\mathcal{L}(\mathcal{F}_{t_{n}} \mid \bb{\psi}) = \prod_{i = 1}^n \left\{ g(t_i \mid \mathcal{F}_{t_{i-1}}; \bb{\zeta})  h(z_i \mid t_i, \mathcal{F}_{t_{i-1}}; \bb{\eta}) f(m_i \mid t_i, z_i, \mathcal{F}_{t_{i-1}}; \bb{\theta}) \right\} \left\{1 - G(T \mid \mathcal{F}_{t_{n}}; {\bb\zeta})\right\} \, ,
\label{exp:spatiolikefacts}
\end{equation}
where $\{z_i\}$ is the collection of random variables corresponding to
the spatial component of the process, which is characterised by the
conditional probability mass or density functions
$h(\cdot \mid t_i, \mathcal{F}_{t_{i-1}}; \bb{\eta})$ with $\bb{\eta}$
being a parameter vector that may or may not share parameters with
$\bb{\zeta}$ and $\bb{\theta}$, and
$\bb{\psi} = (\bb{\zeta}^\top, \bb{\eta}^\top,
\bb{\theta}^\top)^\top$.  The filtration $\mathcal{F}_{t_{i}}$ now
includes all times, marks and locations up to time $t_{i}$.

If the process ends at the last occurrence time $t_n$, then the last
term $1 - G(T \mid \mathcal{F}_{t_{n}}; {\bb\zeta})$
in~(\ref{exp:cox}) and~(\ref{exp:spatiolikefacts}) is not part of the
likelihood \citep[see, for example][Section~4.2]{lindqvist:2006}. This
is the case in the modelling of touch-ball event sequences in football
we consider here, where the process ends with or immediately after the
last event observed in each half of the game.

\section{Bayesian modelling of in-game event sequences}

\subsection{Preamble}

The framework for specifying flexible marked point processes of
Section~\ref{sec:spec} is rather attractive for the modelling of
in-game event sequences in football and other team sports. Firstly,
cross-excitation of in-game events is a natural assumption because any
event in an event sequence is likely to be triggered by one or more of
the previous events. For example, following a corner kick, the next
event is with high probability one among a shot on goal, a defensive
clearance or a claim by the keeper. Such effects can be naturally and
readily captured by the parameters of the conditional mark
distribution in~(\ref{exp:modspecv1}), which involves not only the
magnitudes of all cross-excitations between the various event types
but also the rate at which these excitations decay over
time. Secondly, the preliminary analyses in Section~\ref{sec:lims}
provides strong evidence that occurrence times do not necessarily
cluster, as off-the-shelf marked Hawkes processes imply. Hence, the
freedom to use a more flexible conditional distribution for the
occurrence times, such as a Gamma process, is a rather attractive
prospect. Furthermore, as discussed in Section~\ref{sec:covariates},
team information can be incorporated in the model in a direct way as
covariate information to drive the cross-excitation based on the 
relative abilities of the teams.

Overall, as we demonstrate later, the framework of
Section~\ref{sec:spec} can be used to provide valuable explanatory
tools into the underlying dynamics of the game for the coaching staff
and inform strategic decision making.  It can also produce predictions
of events, such as goals, in a specified time horizon, and of game
outcomes that can enhance, among other things, the viewing experience
in televised games.

\subsection{Excitation-based models}
\label{sec:fittedmodels}

Assume that the touch-ball events in $S$ game periods are $S$
realisations of independent spatio-temporal point processes, with the
$s$th realisation involving $n_s$ events. Denote by $t_{si}$,
$m_{si}$, and $z_{si}$ the occurrence time, mark and location of the
$i$-th event in the $s$th realisation, respectively. Each of the $S$
independent spatio-temporal marked point processes has a likelihood as
in~(\ref{exp:spatiolikefacts}) after dropping the last term, and with
conditional probability mass function for the marks as
in~(\ref{exp:modspecv1}).  The product of the $S$ likelihoods is the
overall likelihood based on the $S$ game periods.

The probability density functions for the occurrence times within each
period are set to
\begin{align}
g(t_{si} \mid \mathcal{F}_{st_{si-1}}; \bb{\zeta}) &= p(t_{si} - t_{si-1} \mid m_{si-1}, \bb{a}, \bb{b}) \nonumber \\
t_{si} - t_{si-1} \mid m_{si-1}, \bb{a}, \bb{b} &\sim \text{Gamma}(a_{m_{si-1}}, b_{m_{si-1}}) \, ,
\label{modspectimes}
\end{align}
where $\mathcal{F}_{st_{si}}$ denotes the filtration of the $s$th
process up to time $t_{si}$.

By this specification, the time to next event is modelled using a
gamma distribution with shape and rate parameters that are specific to
the mark of the last observed event. In this way, we wish to capture
the differences in the expected time to the next event across the
different event types. For example, we expect a shorter time to the
next event after an in-play event like a Pass, compared to
that of an out-of-play event like a Foul. Even within the
group of out-of-play events, we expect a shorter delay following a
Throw-in as compared to a Goal event.

For a discrete set of locations $\{1, \ldots, Z\}$, the conditional
probability mass function for the current location is set to
\begin{align}
  h(z_{si} \mid t_{si}, \mathcal{F}_{st_{si-1}}; \bb{\eta}) & = \eta_{(z_{si-1}, \mathop{m_{si-1}}) \to z_{si}} \, ,
\label{modspeczones}
\end{align}
where $\eta_{(z_{si-1}, \mathop{m_{si-1}}) \to z_{si}}$ is the
probability of transitioning into location $z_{si}$ given the
location $z_{si-1}$ and the mark $m_{si-1}$ of the last observed
event. Expression (\ref{modspeczones}) models the sequence of
locations as a discrete first-order Markov chain \citep[see, for
example,][]{norris1997markov} with a transition probability matrix
$\bb{\eta}$. The current state of the Markov chain is determined by
the combination of the location and the mark of the last observed
event, and the probability of transitioning into the next location
depends only on the current state. The state space of the Markov chain
is given by the Cartesian product
$\{1, \dots, Z\} \times \{1, \dots, M\}$.

We consider four alternative parameterisations for the conditional
probability mass function for the marks. The \SB{} (scalar $\beta$)
spatio-temporal marked point process results
from~(\ref{modspectimes}), (\ref{modspeczones}) and a conditional mark
distribution of the form (\ref{exp:modspecv1}), that is
\begin{equation}
  f(m_{si} \mid t_{si}, \mathcal{F}_{st_{si-1}}; \bb{\theta}) = \frac{\delta_{m_{si}} + \sum_{t_{sj} < t_{si}}
    \mathrm{e}^{\alpha - \beta (t_{si}-t_{sj})}\gamma_{m_{sj} \to m_{si}}}{1 + \sum_{t_{sj} < t_{si}} \mathrm{e}^{\alpha - \beta (t_{si}-t_{sj})}} \, ,
\label{modspecmarksscalarbeta}
\end{equation}
where $\alpha = \log(\alpha^*)$.
The \VB{} (vector $\beta$) model results from~(\ref{modspectimes}),
(\ref{modspeczones}) and
\begin{equation}
f(m_{si} \mid t_i, \mathcal{F}_{st_{si-1}}; \bb{\theta}) =  \frac{\delta_{m_{si}} + \sum_{t_{sj} < t_{si}} \mathrm{e}^{\alpha - \beta_{m_{sj}} (t_{si}-t_{sj})}\gamma_{m_{sj} \to m_{si}}}{1 + \sum_{t_{sj} < t_{si}} \mathrm{e}^{\alpha - \beta_{m_{sj}} (t_{si}-t_{sj})}} \, ,
\label{modspecmarksvectorbeta}
\end{equation}
where $\beta_{m}$ is the exponential decay rate of the excitation
caused by an event of mark $m$. \VB{} allows the decay rates to depend
on the mark of the event causing the excitation. Hence, \SB{} is
formally nested in \VB{} and results when
$\beta = \beta_{1} = \ldots = \beta_{M}$. The \MB{} (matrix $\beta$)
process results from~(\ref{modspectimes}), (\ref{modspeczones}) and
\begin{equation}
f(m_{si} \mid t_{si}, z_{si}, \mathcal{F}_{st_{si-1}}; \bb{\theta}) = \frac{\delta_{m_{si} \mid z_{si}} + \sum_{t_{sj} < t_{si}} \mathrm{e}^{ \alpha - \beta_{m_{sj} \to m_{si} \mid z_{si}} (t_{si}-t_{sj})}\gamma_{m_{sj} \to m_{si} \mid z_{si}}}{ \sum_{m=1}^M \left[ {\delta_{m \mid z_{si}} + \sum_{t_{sj} < t_{si}} \mathrm{e}^{ \alpha - \beta_{m_{sj} \to m \mid z_{si}} (t_{si}-t_{sj})}\gamma_{m_{sj} \to m \mid z_{si}}} \right]} \, ,
\label{modspecmarksmatrix}
\end{equation}
where $\beta_{m \to m' \mid z}$ is the decay rate of the excitation
caused by an event of mark $m$ on an event of mark $m'$ at location
$z$. Specification~(\ref{modspecmarksmatrix}) allows the decay rates
to vary both with the pair of marks involved in the excitation and
across locations, and allows the background mark probabilities
$\bb{\delta}$ and event conversion rates $\bb{\gamma}$ to vary across
location. The \MB{} model can be used to account for scenarios like
those where a \textsf{Corner} event excites a \textsf{Pass} event in
the short term and a \textsf{Shot} event in the longer term
($\beta_{\textsf{Corner} \to \textsf{Pass} \mid 3} >
\beta_{\textsf{Corner} \to \textsf{Shot} \mid 3}$). It also allows us
to capture effects such as how a team is more likely to make more
passes and retain possession of the ball in the defensive zone, while
attempting more shots on goal in the attacking zone
($\gamma_{\textsf{Pass} \to \textsf{Pass} \mid 1} >
\gamma_{\textsf{Pass} \to \textsf{Pass} \mid 3}$ and
$\gamma_{\textsf{Pass} \to \textsf{Shot} \mid 3} >
\gamma_{\textsf{Pass} \to \textsf{Shot} \mid 2}$). The final model we
consider is the \MBA{} (matrix $\beta$ with abilities) where the
baseline category logits of the conversion rates in
(\ref{exp:baselogitspecbayes}) are driven by team information as
\begin{align}
\log\left(\frac{\gamma_{m_{sj} \to m \mid z}(c)}{\gamma_{m_{sj} \to M \mid z}(c)}\right) =  \phi_{m_{sj} \to m \mid z} + \omega_{cm}  \quad (m = 1, \dots, M - 1; c = 1, \ldots, C) \, .
\label{baselogitspecteams}
\end{align}
In the above expression, $\phi_{m \to m' \mid z}$ is a
location-dependent baseline conversion, and $c$ is the team in
possession of the ball attempting the event conversion. The parameter
$\omega_{cm}$, then, reflects the ability of a team to complete a
conversion to an event of mark $m$.

\subsection{Prior distributions}
\label{sec:priorbayes}

The shape and rate parameters of the Gamma distributions for the
inter-arrival times in~(\ref{modspectimes}) are assigned independent
exponential priors with rates $a^{\prime}$ and $b^{\prime}$,
respectively. The probability mass function for the locations
specified in~(\ref{modspeczones}) models the locations as a
multinomial distribution given the current state of the Markov
chain. The conjugate prior for the multinomial distribution is the
Dirichlet distribution \citep[see, for example,][Section
3.4]{gelman2013bayesian} and therefore we assign a Dirichlet prior on
the multinomial probabilities $\bb{\eta}$ with a common concentration
rate parameter $\nu$. The background mark probability vector
$\bb{\delta}$ in the \SB{}, \VB{}, \MB{} and \MBA{} models is also
assigned a Dirichlet prior with concentration hyper-parameter
$\delta^{\prime}$. The location-specific mark probability vectors
$(\delta_{1 \mid 1}, \ldots, \delta_{M \mid 1})^\top, \ldots,
(\delta_{1 \mid Z}, \ldots, \delta_{M \mid Z})^\top$ in the \MB{} and
\MBA{} models are assigned independent Dirichlet priors with
concentration hyper-parameter $\delta^{\prime\prime}$. The excitation
factor $\alpha$ is assigned a normal prior with mean $0$ and standard
deviation $\sigma_\alpha$. The decay rate parameter $\beta$ in the
\SB{} model, the parameters $\beta_1, \ldots, \beta_M$ in the \VB{}
model, and their location-specific counterparts in the \MB{} and
\MBA{} models are assigned independent exponential priors with a
common rate $\beta^{\prime}$. The parameters $\phi_{m \to m' \mid z}$
and $\omega_{cm}$
$(m, m' = 1, \ldots, M; z = 1, \ldots, Z; c = 1, \ldots, C)$ in the
\MBA{} model are assigned independent Normal priors with mean $0$ and
standard error $\sigma_{\gamma}$. The conversion rate parameters
$(\gamma_{m \to 1}, \ldots, \gamma_{m \to M})^\top$ in the \SB{} and
\VB{} models, and their location-specific counterparts in the \MB{}
model are assigned independent Dirichlet priors with a common
concentration rate parameter $\gamma^\prime$.

\subsection{Posterior distributions}

The time and location conditional distributions corresponding to
(\ref{modspectimes}) and (\ref{modspeczones}) share no parameters with
each other, and no parameters with any of the conditional mark
distributions for the \SB{}, \VB{}, \MB{}, and \MBA{}
models. Furthermore, the likelihood in~(\ref{exp:spatiolikefacts}) can
be factorised into a term depending only on the time parameters
$\bb{\zeta}$, a term depending only on the location parameters
$\bb{\eta}$ and a term depending only on the mark parameters
$\bb{\theta}$. Given that the priors for $\bb{\zeta}$, $\bb{\eta}$ and
$\bb{\theta}$ are also independent, the derivation of, or sampling from,
the posterior distributions can be performed separately for each of
those parameters.

The priors for the location parameters $\bb{\eta}$ are conjugate, so
the posterior for $\bb{\eta}$ is readily obtained. If
$\bb{y} = \{y_{i \to j}\}$, for $j \in \{1, \dots, Z\}$, are the
observed counts of transitions originating from the state $i$ where
$i \in \{1, \dots, Z\} \times \{1, \dots, M\}$, then the posterior
distribution of each row of the transition matrix $\bb{\eta_i}$ is a
Dirichlet distribution with concentration parameters
$(y_{i1} + \nu, \ldots, y_{iZ} + \nu)$.

Posterior sampling for the parameter vectors $\bb{a}$, $\bb{b}$
in~(\ref{modspectimes}) of the conditional distributions for the
times, and the parameters $\bb{\theta}$ of the conditional
distributions for the marks in each of the \SB{}, \VB{}, \MB{}, and
\MBA{} models is carried out using the variant of the Hamiltonian
Monte Carlo procedure \citep{duane1987hybrid} that is implemented in
\textsf{Stan} \citep{cmdstan}.

We have also implemented posterior sampling using a
Metropolis-within-Gibbs procedure, which, though, proved to mix poorly
in artificial data sets for the \SB{}, \VB{}, \MB{}, and \MBA{},
rendering it computationally infeasible. As in the case of Hawkes process,
the poor mixing stems from the presence of strong correlations between 
the model parameters as well as the flatness of the likelihood function 
\citep{veen2008estimation}. \textsf{Stan}, on the other hand, 
implements the No-U-Turn Sampler \citep{hoffman2014no} that 
automatically calibrates tuning parameters in a warm-up phase and 
can efficiently sample from complex posterior distributions.

\subsection{Model complexity}
\label{sec:assocrules}

The conditional mark distribution in the \MB{} and \MBA{} models 
involves a large number of parameters. There are $M^2 Z$ decay rate 
parameters and $M(M - 1)Z$ baseline conversion rate parameters 
which makes posterior sampling a computationally challenging task. 
We have developed a screening procedure based on association rule 
learning \citep[see, for example,][]{agarwal1993mining} that operates 
on the data involved in the likelihood and eliminates parameters prior 
to posterior sampling.

The screening procedure retains only those parameters that capture the
most significant event interactions and depends on two constants that
need to be chosen. The first is the window size $W$ for the number of
transient events, and any event is allowed to be triggered by only one
of the $W$ events leading up to it. The other is the number of event
pairs $N$ considered in each of the three zones and sets a threshold
on the number of significant event interactions that are
identified. Full details on the association rule based screening
procedure are given in Section~S2 of the Supporting Materials.

\subsection{Model evaluation}
\label{sec:lpd}

Let ${\bb X}^{(train)}$ be the set of the training data on which the
likelihood is based on, consisting of $n^{(train)}$ events, and let
$\bb{\psi}^{(1)}, \ldots, \bb{\psi}^{(R)}$ be $R$ samples from the
posterior distribution. Denote by ${\bb X}^{(test)}$ the set of
held-out test data, consisting of $n^{(test)}$ events.
  
One method to evaluate the predictive accuracy of each model is to use
the log point-wise predictive density \citep{vehtari2017practical}
computed on the test data, using the posterior samples
\begin{equation}
  \label{exp:logpointwise}
  \widehat{lpd} = \sum_{(t, z, m) \in {\bb X}^{(test)}} \log \left( \frac{1}{R} \sum_{r = 1}^R L(t, z, m  \mid \mathcal{F}_{t^-}, \bb{\zeta}^{(r)}, \bb{\eta}^{(r)}, \bb{\theta}^{(r)})\right) \, ,
\end{equation}
where
$L(t, z, m \mid \mathcal{F}_{t^-}, \bb{\zeta}^{(r)}, \bb{\eta}^{(r)},
\bb{\theta}^{(r)})$ is the likelihood of $(t, z, m)$ given the
filtration $\mathcal{F}_{t^-}$ at the posterior sample
$\bb{\zeta}^{(r)}, \bb{\eta}^{(r)}, \bb{\theta}^{(r)}$. Large values of 
$\widehat{lpd}$ indicate better predictive accuracy.

Apart from \SB{}, \VB{}, \MB{}, and \MBA{}, we also evaluate the
predictive accuracy of two simpler baseline models that do not include 
Hawkes-like excitation effects. The first baseline model, termed \FOMC{} 
model, is based on the factorisation of the likelihood of marked 
spatio-temporal processes in expression~(\ref{exp:spatiolikefacts}) with 
models for the times and locations as in~(\ref{modspectimes}) and
(\ref{modspeczones}), respectively, but with the conditional probability
mass function for the marks being a first-order Markov chain
\begin{equation}
  f(m_i \mid t_i, z_i, \mathcal{F}_{t_{i-1}}; \bb{\theta}) = \theta_{(z_{i}, \mathop{m_{i-1}}) \to m_{i}} \, .
\label{modspecmarksmarkov}
\end{equation}
In this specification, $\theta_{(z, \mathop{m}) \to m'}$ is the
probability of the event mark $m'$ given that last observed event has
location $z$ and mark $m$. The second baseline model, termed \MSTHP{}, 
is the marked spatio-temporal homogeneous Poisson process
(\citealt[Section 7.3]{daley2003theory}), which has likelihood
\begin{equation}
  \label{exp:poissonlik}
  \mathcal{L}^{(P)}(\bb{q} \mid \bb{\rho}) = \prod_{m = 1}^{M} \prod_{z = 1}^{Z} \rho_{mz}^{q_{mz}} \exp\left\{ -T \rho_{mz}  \right\} \, ,
\end{equation}
where $\rho_{mz}$ is the Poisson rate parameter and $q_{mz}$ is the
number of event occurrences for mark $m$ at location $z$ over a total
observation time $T$ in the data.

The \FOMC{} and \MSTHP{} models have conjugate prior distributions and 
therefore their posteriors are readily obtained. Details on those prior 
distributions and the derivation of their posterior distributions are given 
in Section~S1 of the Supporting Materials.

\begin{table}[t]
\centering
\caption{Zone-wise frequencies for each event type in the training
  data.}
 \resizebox{0.8\linewidth}{!}{
\begin{tabular}[t]{clrrr}
  \toprule
\multicolumn{5}{c}{Home} \\
\multicolumn{2}{c}{mark} & \multicolumn{3}{c}{zone} \\
\cmidrule(l{3pt}r{3pt}){1-2} \cmidrule(l{3pt}r{3pt}){3-5}
m & label & 1 & 2 & 3\\
\midrule
1 & Home\_Win & 236 & 257 & 41\\
2 & Home\_Dribble & 17 & 96 & 96\\
3 & Home\_Pass\_S & 1699 & 4633 & 1658\\
4 & Home\_Pass\_U & 541 & 825 & 725\\
5 & Home\_Shot & 0 & 0 & 292\\
6 & Home\_Keeper & 155 & 0 & 0\\
7 & Home\_Save & 106 & 0 & 0\\
8 & Home\_Clear & 557 & 141 & 32\\
9 & Home\_Lose & 122 & 287 & 368\\
\addlinespace
10 & Home\_Goal & 0 & 0 & 22\\
11 & Home\_Foul & 62 & 126 & 64\\
12 & Home\_Out\_Throw & 97 & 184 & 163\\
13 & Home\_Out\_GK & 149 & 0 & 0\\
14 & Home\_Out\_Corner & 0 & 0 & 112\\
15 & Home\_Pass\_O & 7 & 19 & 20\\ \bottomrule
\end{tabular}
\begin{tabular}[t]{clrrr}
  \toprule
  \multicolumn{5}{c}{Away} \\
\multicolumn{2}{c}{mark} & \multicolumn{3}{c}{zone} \\
\cmidrule(l{3pt}r{3pt}){1-2} \cmidrule(l{3pt}r{3pt}){3-5}
m & label & 1 & 2 & 3\\ \midrule
16 & Away\_Win & 25 & 204 & 301\\
17 & Away\_Dribble & 76 & 65 & 19\\
18 & Away\_Pass\_S & 1427 & 4390 & 2030\\
19 & Away\_Pass\_U & 542 & 811 & 702\\
20 & Away\_Shot & 193 & 2 & 0\\
21 & Away\_Keeper & 0 & 0 & 192\\
22 & Away\_Save & 0 & 0 & 149\\
23 & Away\_Clear & 27 & 124 & 660\\
24 & Away\_Lose & 323 & 349 & 142\\
\addlinespace
25 & Away\_Goal & 20 & 0 & 0\\
26 & Away\_Foul & 46 & 112 & 69\\
27 & Away\_Out\_Throw & 143 & 173 & 110\\
28 & Away\_Out\_GK & 0 & 0 & 220\\
29 & Away\_Out\_Corner & 76 & 0 & 0\\
30 & Away\_Pass\_O & 13 & 11 & 5\\
\bottomrule
\end{tabular}}
\label{trainingtable}
\end{table}

\section{Explanatory modelling}

\begin{table}[t]
\centering
\caption{\label{tab:matrixsummary}Posterior summaries and convergence diagnostics from 2000 posterior samples for selected parameters from the \MBA{} model after screening with $(W, N) = (5, 100)$.}
\resizebox{0.8\linewidth}{!}{
\begin{tabular}[t]{lrrrrr}
\toprule
parameter & mean & sd & $\hat{R}$ & $N^{(eff)}$\\ 
\midrule
$\beta_{3 \to 3 \mid 1}$  & 0.52 & 0.04 & 1.00 & 1309.81\\
$\beta_{27 \to 8 \mid 1}$ & 1.97 & 0.86 & 1.00 & 1953.81\\
$\beta_{24 \to 1 \mid 2}$ & 1.51 & 0.09 & 1.00 & 2042.84\\
$\beta_{3 \to 4 \mid 2}$  & 0.65 & 0.03 & 1.01 & 913.52\\
$\beta_{3 \to 5 \mid 3}$  & 0.63 & 0.04 & 1.01 & 1933.43\\
$\beta_{3 \to 10 \mid 3}$ & 0.81 & 0.24 & 1.01 & 882.31\\
$\phi_{3 \to 3 \mid 1}$   & 1.70 & 0.50 & 1.01 & 792.46\\
$\phi_{27 \to 8 \mid 1}$  & -0.74 & 0.87 & 1.00 & 2159.96\\
$\phi_{24 \to 1 \mid 2}$  & 3.29 & 0.37 & 1.02 & 539.25\\
\bottomrule
\end{tabular}
\begin{tabular}[t]{lrrrrr}
\toprule
parameter & mean & sd & $\hat{R}$ & $N^{(eff)}$\\ 
\midrule
$\phi_{3 \to 4 \mid 2}$   & 1.81 & 0.41 & 1.01 & 540.41\\
$\phi_{3 \to 5 \mid 3}$   & 1.38 & 0.35 & 1.01 & 576.16\\
$\phi_{3 \to 10 \mid 3}$  & -1.31 & 0.59 & 1.01 & 1098.83\\
$\delta_{3 \mid 1}$       & 0.56 & 0.03 & 1.00 & 1805.80\\
$\delta_{3 \mid 2}$       & 0.24 & 0.08 & 1.00 & 1356.85\\
$\delta_{3 \mid 3}$       & 0.03 & 0.01 & 1.00 & 2207.39\\
$\alpha$                  & 6.30 & 0.09 & 1.01 & 866.96\\
$\omega_{9,3}$            & 0.59 & 0.26 & 1.03 & 334.15\\
$\omega_{10,3}$           & 0.67 & 0.25 & 1.01 & 341.24\\
\bottomrule
\end{tabular}}
\end{table}

\subsection{Training\label{sec:modtrain}}
Samples from the posterior distributions for the parameters of the
\SB{}, \VB{}, \MB{}, and \MBA{} models of
Section~\ref{sec:fittedmodels}, and of the \FOMC{} and \MSTHP{}
baseline models of Section~\ref{sec:lpd} are obtained using all event
sequences from the first $20$ games of the league season, played
between 17/08/2013 and 26/08/2013, which constitute
$\bb{X}^{(train)}$. The training data involves $S = 40$ game periods
involving of $27,660$ events. Each of the $20$ teams participating in
the league plays in two of the $20$ games, one at their home and one
at their away venue. As is also described in
Section~\ref{sec:descriptives}, there are $M = 30$ marks and $Z = 3$
zones. Table~\ref{trainingtable} gives the zone-wise event frequencies
across marks in the training data used for modelling, reflecting the
large variability in the frequencies both across marks and zones.

For the \MB{} and \MBA{} models, the association rule learning
screening procedure of Section~\ref{sec:assocrules} is employed for
all combinations of $W \in \{5, 10\}$ and $N \in \{50, 100\}$ to
eliminate some of the model parameters and reduce the model complexity
before posterior sampling.

The hyper-parameters for the prior distributions specified in
Section~\ref{sec:priorbayes} are as follows. The hyper-parameters
$a^{\prime}$ and $b^{\prime}$ for the exponential priors on the
parameters of the Gamma distributions in~(\ref{modspectimes}) are both
set to $0.01$. The Dirichlet prior on the background mark
probabilities $\bb{\delta}$ has concentration hyper-parameter
$\delta^{\prime} = 1$. The exponential prior on the decay rates
$\bb{\beta}$ has a rate hyper-parameter $\beta^{\prime} = 0.1$. The
Normal priors on the excitation factor $\alpha$ and the
baseline-category logit model parameters $\bb{\phi}$ and $\bb{\omega}$
have hyper-parameters $\sigma_{\alpha}, \sigma_{\gamma} = 10$. The 
location-specific background mark probability vectors in the \MB{} and
\MBA{} models are assigned independent Dirichlet priors with
concentration hyper-parameter $\delta^{\prime\prime} = 1$.

The ability parameters in the baseline category logit specification
for the conversion rates of the \MBA{} are not directly
identifiable. In order to make them so, we set the abilities
$\omega_{cm}$ for West Ham United to $0$ $(m = 1, \ldots, 30)$. 
Then, $\omega_{cm} > 0$ indicates that for team $c$, a
previous event is more likely to trigger an event of mark $m$ when
compared to the reference team.

Samples from the posterior distributions are obtained by running four
parallel chains using the Hamiltonian Monte Carlo procedures
implemented in \textsf{Stan}. The \textsf{Stan} templates we used are
all provided in the Supporting Materials. Each chain 
is initialised with different starting values and run for a total of 500 
iterations post the warm-up phase. Table~\ref{tab:matrixsummary} 
gives posterior summaries along with convergence diagnostics for some 
of the parameters of the \MBA{} model with $W = 5$ and $N = 100$; the 
corresponding chain-wise trace plots are provided in the Supporting 
Materials. 

\begin{figure}[!t]
\centering
\resizebox{\linewidth}{!}{
\includegraphics{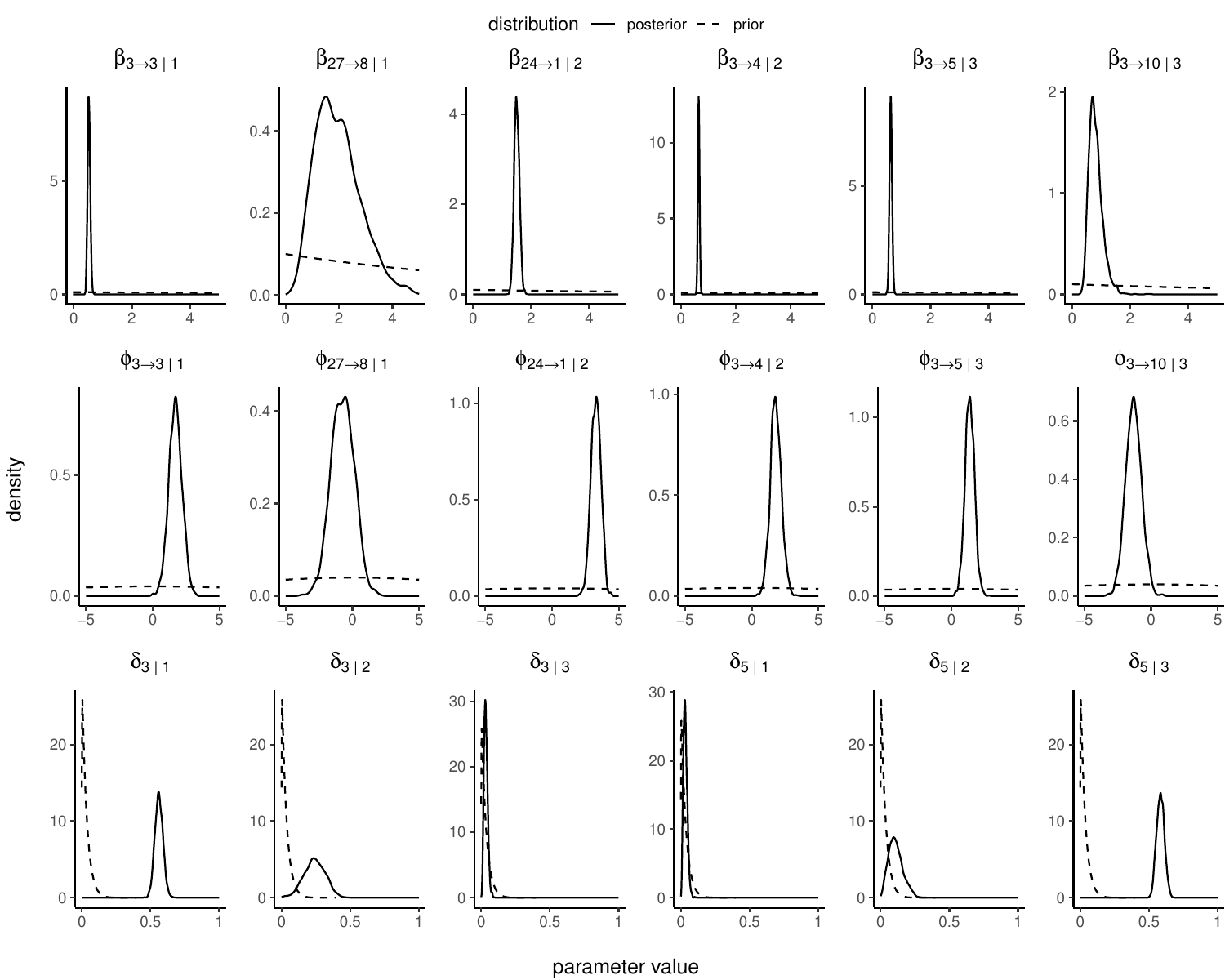}}
\caption{Visualising the impact of prior specifications by overlaying
  the posterior and prior densities for selected model parameters for
  the \MBA{} model with $(W, N) = (5, 100)$.}
\label{fig:postprioroverlay}
\end{figure}
  
The convergence of the algorithm is assessed using the potential scale
reduction factor $\hat{R}$ proposed by \citet{gelman1992inference},
which is the ratio of the average variance within each chain to the
variance of the aggregated samples across chains. If the chains have
converged to the stationary distribution, the expected value of
$\hat{R}$ is 1.  All parameters have $\hat{R} < 1.1$, which, as
recommended in \citet{gelman1992inference}, is evidence for
convergence. Table~\ref{tab:matrixsummary} also gives the effective
sample size \citep[see, for example,][Section
11.5]{gelman2013bayesian} for the samples for each of the posterior
marginals, which indicate that the sampler returned samples with
acceptable autocorrelation. For some parameters the effective sample
size is larger than the sample size due to negative
autocorrelations. This, as pointed out in \citet{vehtari2021rank}, is
a consequence of the HMC algorithm used in \textsf{Stan} being an
antithetic Markov chain which has negative autocorrelations on odd
lags. The impact of the prior distributions in
Section~\ref{sec:priorbayes} on the posterior samples is minimal as
seen, for a selection of parameters, in Figure
\ref{fig:postprioroverlay} indicating that the posterior distributions
of the parameters have concentrated after accounting for the
likelihood.

\begin{table}[t]
\centering
\caption{\label{tab:logpostdensity}Cumulative log posterior densities $\widehat{lpd}$ over 10 game periods in the test data for all fitted models along with the number of estimated parameters $d^{(par)}$ in each model. For the \MB{} models, $W$ is the number of transient events and $N$ is the number of significant event pairs identified in the rule-based framework for reducing model complexity.}
\resizebox{0.8\linewidth}{!}{
\begin{tabular}[t]{llrr}
\toprule
model & abbreviation & $d^{(par)}$ & $\widehat{lpd}$\\
\midrule
Homogeneous Poisson process (Baseline) & \MSTHP{} & 90 & -35469.50\\
Matrix $\beta$ $(W, N) = (10, 50)$ & \MB{2} & 538 & -22288.64\\
Matrix $\beta$ $(W, N) = (5, 50)$ & \MB{1} & 538 & -22152.08\\
First order Markov chain (Baseline) & \FOMC{} & 870 & -21898.31\\
Scalar $\beta$ & \SB{} & 902 & -21838.04\\
Vector $\beta$ & \VB{} & 915 & -21829.81\\
Matrix $\beta$ with abilities $(W, N) = (5, 100)$ & \MBA{} & 1539 & -21599.81\\
Matrix $\beta$ $(W, N) = (10, 100)$ & \MB{4} & 988 & -21496.56\\
Matrix $\beta$ $(W, N) = (5, 100)$ & \MB{3} & 988 & -21342.57\\
\bottomrule
\end{tabular}}
\end{table}

\subsection{Model evaluation\label{sec:modeval}}

The \SB{}, \VB{}, \MB{}, and \MBA{} models are compared with each
other and with the \FOMC{} and \MSTHP{} baseline models of
Section~\ref{sec:lpd} in terms of their log point-wise predictive
density~(\ref{exp:logpointwise}) computed on test data. The test data
$\bb{X}^{(test)}$ includes all events from the 5 games
immediately following the games in the training data, played between 
31/08/2013 and 01/09/2013. The test data involves $S = 10$
game periods involving of $27,660$ events.

Table~\ref{tab:logpostdensity} gives the log predictive densities
$\widehat{lpd}$, summed over all the events in the $10$
game periods in the test data. Table~\ref{tab:logpostdensity} also
provides the number of parameters in each model as a measure of their
complexity. The three top performing models are the \MB{} model after
screening with $(W, N) = (5, 100)$, followed by \MB{} model after
screening with $(W, N) = (10, 100)$, and \MBA{} with
$(W, N) = (5, 100)$. Notably, the \MB{} model after screening with
$(W, N) = (5, 100)$ performs the best among the list of fitted models,
significantly outperforming also models of similar complexity, such as
the \SB{}, \VB{} and \FOMC{} models. The slightly poorer performance
of the \MBA{} model with $(W, N) = (5, 100)$ is most probably due to
the fact that, in the training data, each team plays just one game at
their home and one at their away venue. Nevertheless, in order to
illustrate the full explanatory potential of the modelling framework
in Section~\ref{sec:spec}, we focus on inferences based on the
posterior samples from the \MBA{} model.

\subsection{Background mark probabilities}

\begin{table}[t]
\centering
\caption{\label{tab:deltameans}Posterior means and standard deviations (in  parenthesis) of the zone dependent background mark probabilities $\delta_{m \mid z}$ for $z \in \{1, 2, 3\}$ from the \MBA{} model. The dots ($\cdot$) denote means and standard deviations less than 0.005.}
\resizebox{0.8\linewidth}{!}{
\begin{tabular}[t]{p{3.25cm}p{1.5cm}p{1.5cm}p{1.5cm}}
\toprule
mark label & 1 & 2 & 3 \\ 
\midrule
Home\_Win & . & . & .\\\\
Home\_Dribble & . & 0.01 (0.01) & .\\
Home\_Pass\_S & 0.56 (0.03) & 0.24 (0.08) & 0.03 (0.01)\\
Home\_Pass\_U & 0.06 (0.01) & 0.04 (0.02) & 0.03 (0.01)\\
Home\_Shot & . & . & .\\\\
Home\_Keeper & 0.04 (0.01) & . & .\\
Home\_Save & . & . & .\\\\
Home\_Clear & 0.05 (0.02) & . & .\\
Home\_Lose & . & 0.04 (0.02) & .\\
Home\_Goal & . & . & .\\\\
Home\_Foul & 0.03 (0.01) & 0.09 (0.04) & 0.02 (0.01)\\
Home\_Out\_Throw & . & . & .\\\\
Home\_Out\_GK & . & . & .\\\\
Home\_Out\_Corner & . & . & .\\\\
Home\_Pass\_O & . & 0.08 (0.02) & .\\
\bottomrule
\end{tabular}
\begin{tabular}[t]{p{3.25cm}p{1.5cm}p{1.5cm}p{1.5cm}}
\toprule
mark label & 1 & 2 & 3 \\ 
\midrule
Away\_Win & . & . & .\\\\
Away\_Dribble & . & 0.02 (0.02) & .\\
Away\_Pass\_S & 0.03 (0.01) & 0.11 (0.05) & 0.58 (0.03)\\
Away\_Pass\_U & 0.02 (0.01) & 0.07 (0.04) & 0.05 (0.01)\\
Away\_Shot & . & . & .\\\\
Away\_Keeper & . & . & 0.05 (0.01)\\
Away\_Save & . & . & .\\\\
Away\_Clear & . & . & .\\\\
Away\_Lose & 0.02 (0.01) & 0.07 (0.03) & 0.06 (0.01)\\
Away\_Goal & . & . & .\\\\
Away\_Foul & 0.02 (0.01) & 0.06 (0.03) & .\\
Away\_Out\_Throw & . & . & .\\\\
Away\_Out\_GK & . & . & .\\\\
Away\_Out\_Corner & . & . & .\\\\
Away\_Pass\_O & . & 0.05 (0.02) & .\\
\bottomrule
\end{tabular}}
\end{table}

Table~\ref{tab:deltameans} gives the posterior means of the background
probabilities $\delta_{m \mid z}$ for all marks
$m \in \{1, \ldots, 30\}$ and all locations $z \in \{1, 2, 3\}$. The
background mark probabilities for the home and away team events in 
zone 1 are almost equal to those in zone 3 for the away and home team 
events, respectively. This is as expected because the attacking zone for 
the home team is the defensive zone for the away team and vice-versa.

The similar probabilities for the home and away background mark
probabilities could indicate that the background process of the game is 
not influenced by home advantage. To confirm this, we fit another 
\MBA{} model after constraining all the corresponding home and away 
background mark probabilities to be equal, for example, 
$\delta_{\textsf{Home\_Pass\_S} \mid 1} = \delta_{\textsf{Away\_Pass\_S} 
\mid 3}$, $\delta_{\textsf{Home\_Foul} \mid 3} = 
\delta_{\textsf{Away\_Foul} \mid 1}$, $\delta_{\textsf{Home\_Dribble} 
\mid 2} = \delta_{\textsf{Away\_Dribble} \mid 2}$ and so on. The 
constrained \MBA{} model has 45 fewer parameters to be estimated 
as compared to the full \MBA{} model.

\begin{table}[!t]
\caption{\label{tab:gamma_zone}Posterior means and standard 
deviations (in parenthesis) of the event conversion probabilities 
$\gamma_{m_{j} \to m_{i} \mid z_i}$ corresponding to the location 
$z = 2$ from the \MBA{} model for 
Manchester united. The $\gamma_{m \to m' \mid z}$'s are computed 
by setting the team identifier $c = 11$, (corresponding to Manchester 
United), for both the home as well as the away events. The highlighted 
cells (in bold) illustrate the superior ability of Manchester United 
when playing at home compared to away. The dots ($\cdot$) denote means and standard deviations less than 0.005.}
\centering
\resizebox{0.975\linewidth}{!}{
\begin{tabular}[t]{lp{0.9cm}p{0.5cm}p{0.9cm}p{0.9cm}p{0.5cm}p{0.5cm}p{0.5cm}p{0.9cm}p{0.9cm}p{0.5cm}p{0.9cm}p{0.9cm}p{0.5cm}p{0.5cm}p{0.5cm}p{0.9cm}p{0.5cm}p{0.9cm}p{0.9cm}p{0.5cm}p{0.5cm}p{0.5cm}p{0.9cm}p{0.9cm}p{0.5cm}p{0.9cm}p{0.9cm}p{0.5cm}p{0.5cm}p{0.9cm}}
\toprule
\rotatebox{90}{ } & \rotatebox{90}{Home\_Win} & \rotatebox{90}{Home\_Dribble} & \rotatebox{90}{Home\_Pass\_S} & \rotatebox{90}{Home\_Pass\_U} & \rotatebox{90}{Home\_Shot} & \rotatebox{90}{Home\_Keeper} & \rotatebox{90}{Home\_Save} & \rotatebox{90}{Home\_Clear} & \rotatebox{90}{Home\_Lose} & \rotatebox{90}{Home\_Goal} & \rotatebox{90}{Home\_Foul} & \rotatebox{90}{Home\_Out\_Throw} & \rotatebox{90}{Home\_Out\_GK} & \rotatebox{90}{Home\_Out\_Corner} & \rotatebox{90}{Home\_Pass\_O} & \rotatebox{90}{Away\_Win} & \rotatebox{90}{Away\_Dribble} & \rotatebox{90}{Away\_Pass\_S} & \rotatebox{90}{Away\_Pass\_U} & \rotatebox{90}{Away\_Shot} & \rotatebox{90}{Away\_Keeper} & \rotatebox{90}{Away\_Save} & \rotatebox{90}{Away\_Clear} & \rotatebox{90}{Away\_Lose} & \rotatebox{90}{Away\_Goal} & \rotatebox{90}{Away\_Foul} & \rotatebox{90}{Away\_Out\_Throw} & \rotatebox{90}{Away\_Out\_GK} & \rotatebox{90}{Away\_Out\_Corner} & \rotatebox{90}{Away\_Pass\_O}\\
\midrule
Home\_Win & . & . & \textbf{0.41 (0.1)} & 0.03 (0.02) & . & . & . & . & 0.01 (0.01) & . & . & . & . & . & . & 0.01 (0.02) & . & . & 0.03 (0.02) & . & . & . & 0.02 (0.02) & . & . & . & 0.44 (0.11) & . & . & .\\
Home\_Dribble & . & . & \textbf{0.72 (0.14)} & 0.07 (0.05) & . & . & . & . & 0.05 (0.03) & . & . & . & . & . & . & 0.07 (0.09) & . & . & . & . & . & . & . & . & . & . & . & . & . & 0.08 (0.05)\\
Home\_Pass\_S & . & . & \textbf{0.8 (0.06)} & 0.08 (0.01) & . & . & . & . & . & . & . & . & . & . & . & . & . & 0.03 (0.04) & . & . & . & . & . & . & . & . & 0.02 (0.03) & . & . & .\\
Home\_Pass\_U & 0.02 (0.02) & . & 0.02 (0.02) & 0.01 (0.01) & . & . & . & 0.02 (0.02) & . & . & . & 0.02 (0.02) & . & . & . & 0.17 (0.04) & . & 0.22 (0.03) & 0.07 (0.01) & . & . & . & 0.15 (0.03) & . & . & . & 0.23 (0.05) & . & . & .\\
Home\_Shot & . & . & 0.49 (0.24) & . & . & . & . & . & . & . & . & . & . & . & . & . & . & 0.31 (0.21) & . & . & . & . & . & . & . & . & . & . & . & 0.2 (0.12)\\
Home\_Keeper & . & . & 0.4 (0.23) & 0.15 (0.14) & . & . & . & . & . & . & . & . & . & . & . & . & . & . & 0.15 (0.15) & . & . & . & 0.17 (0.16) & . & . & . & . & . & . & 0.13 (0.07)\\
Home\_Save & . & . & 0.4 (0.23) & . & . & . & . & . & . & . & . & . & . & . & . & . & . & 0.4 (0.23) & . & . & . & . & . & . & . & . & . & . & . & 0.2 (0.12)\\
Home\_Clear & . & . & 0.04 (0.03) & 0.02 (0.01) & . & . & . & . & . & . & . & 0.04 (0.03) & . & . & . & 0.02 (0.01) & . & 0.13 (0.03) & 0.04 (0.02) & . & . & . & . & 0.02 (0.02) & . & . & 0.64 (0.09) & . & . & .\\
Home\_Lose & 0.02 (0.02) & . & 0.02 (0.02) & 0.01 (0.01) & . & . & . & 0.02 (0.02) & . & . & . & 0.02 (0.02) & . & . & . & 0.59 (0.08) & . & 0.07 (0.02) & . & . & . & . & . & 0.1 (0.03) & . & . & 0.1 (0.04) & . & . & .\\
Home\_Goal & . & . & . & . & . & . & . & . & . & . & . & . & . & . & . & . & . & 0.55 (0.22) & . & . & . & . & . & . & . & . & . & . & . & 0.45 (0.22)\\
Home\_Foul & . & . & 0.48 (0.15) & 0.24 (0.11) & . & . & . & . & . & . & . & . & . & . & . & . & . & . & 0.09 (0.09) & . & . & . & . & . & . & . & . & . & . & 0.19 (0.1)\\
Home\_Out\_Throw & . & . & 0.86 (0.05) & 0.09 (0.03) & . & . & . & . & . & . & . & . & . & . & . & . & . & . & . & . & . & . & . & . & . & . & . & . & . & .\\
Home\_Out\_GK & . & . & . & 0.06 (0.05) & . & . & . & . & 0.07 (0.09) & . & 0.04 (0.05) & 0.25 (0.2) & . & . & . & . & . & . & 0.14 (0.13) & . & . & . & 0.09 (0.06) & 0.07 (0.08) & . & . & 0.2 (0.18) & . & . & 0.07 (0.03)\\
Home\_Out\_Corner & . & . & 0.7 (0.18) & . & . & . & . & . & . & . & . & . & . & . & . & . & . & . & . & . & . & . & . & . & . & . & . & . & . & 0.3 (0.18)\\
Home\_Pass\_O & . & . & . & . & . & . & . & . & . & . & . & . & . & . & . & . & . & 0.4 (0.19) & 0.25 (0.15) & . & . & . & . & . & . & . & . & . & . & 0.35 (0.16)\\
\addlinespace
Away\_Win & 0.05 (0.04) & . & . & 0.01 (0.01) & . & . & . & 0.02 (0.02) & 0.03 (0.02) & . & . & 0.59 (0.1) & . & . & . & . & . & \textbf{0.22 (0.07)} & 0.02 (0.01) & . & . & . & . & . & . & . & . & . & . & 0.03 (0.01)\\
Away\_Dribble & 0.31 (0.24) & . & . & . & . & . & . & . & . & . & . & . & . & . & . & . & . & \textbf{0.49 (0.23)} & 0.06 (0.05) & . & . & . & . & 0.03 (0.03) & . & . & . & . & . & 0.11 (0.06)\\
Away\_Pass\_S & 0.03 (0.04) & . & 0.06 (0.09) & 0.02 (0.02) & . & . & . & 0.02 (0.03) & 0.02 (0.03) & . & . & 0.06 (0.09) & . & . & . & . & . & \textbf{0.68 (0.11)} & 0.06 (0.01) & . & . & . & . & . & . & . & . & . & . & .\\
Away\_Pass\_U & 0.25 (0.05) & . & 0.2 (0.03) & 0.07 (0.01) & . & . & . & 0.14 (0.03) & . & . & . & 0.28 (0.05) & . & . & . & . & . & . & . & . & . & . & . & . & . & . & . & . & . & .\\
Away\_Shot & . & . & 0.44 (0.22) & . & . & . & . & . & . & . & . & . & . & . & . & . & . & 0.36 (0.21) & . & . & . & . & . & . & . & . & . & . & . & 0.2 (0.12)\\
Away\_Keeper & . & . & . & 0.16 (0.15) & . & . & . & 0.19 (0.16) & . & . & . & . & . & . & . & . & . & 0.42 (0.22) & 0.11 (0.11) & . & . & . & . & . & . & . & . & . & . & 0.13 (0.07)\\
Away\_Save & . & . & 0.47 (0.22) & . & . & . & . & . & . & . & . & . & . & . & . & . & . & 0.31 (0.2) & . & . & . & . & . & . & . & . & . & . & . & 0.22 (0.13)\\
Away\_Clear & 0.02 (0.01) & . & 0.15 (0.04) & 0.05 (0.02) & . & . & . & 0.02 (0.01) & . & . & . & 0.69 (0.07) & . & . & . & . & . & 0.03 (0.02) & . & . & . & . & . & . & . & . & 0.01 (0.01) & . & . & .\\
Away\_Lose & 0.69 (0.06) & . & 0.05 (0.01) & . & . & . & . & . & 0.08 (0.02) & . & . & 0.1 (0.03) & . & . & . & . & . & . & . & . & . & . & . & . & . & . & . & . & . & .\\
Away\_Goal & . & . & 0.73 (0.17) & . & . & . & . & . & . & . & . & . & . & . & . & . & . & . & . & . & . & . & . & . & . & . & . & . & . & 0.27 (0.17)\\
Away\_Foul & . & . & . & 0.13 (0.12) & . & . & . & . & . & . & . & . & . & . & . & . & . & 0.5 (0.15) & 0.18 (0.09) & . & . & . & . & . & . & . & . & . & . & 0.18 (0.1)\\
Away\_Out\_Throw & 0.02 (0.02) & . & 0.02 (0.02) & . & . & . & . & 0.01 (0.01) & . & . & . & 0.01 (0.02) & . & . & . & . & . & 0.86 (0.06) & 0.05 (0.02) & . & . & . & . & . & . & . & . & . & . & .\\
Away\_Out\_GK & . & . & . & 0.09 (0.08) & . & . & . & 0.14 (0.08) & 0.13 (0.12) & . & . & 0.19 (0.14) & . & . & . & . & . & . & 0.11 (0.08) & . & . & . & . & 0.1 (0.09) & . & 0.06 (0.06) & 0.1 (0.09) & . & . & 0.08 (0.03)\\
Away\_Out\_Corner & . & . & . & . & . & . & . & . & . & . & . & . & . & . & . & . & . & 0.6 (0.2) & . & . & . & . & . & . & . & . & . & . & . & 0.4 (0.2)\\
Away\_Pass\_O & . & . & 0.74 (0.15) & 0.1 (0.08) & . & . & . & . & . & . & . & . & . & . & . & . & . & . & . & . & . & . & . & . & . & . & . & . & . & 0.16 (0.11)\\
\bottomrule
\end{tabular}}
\end{table}

The formal method to test our hypothesis is to calculate the Bayes
factor, defined as the ratio of the marginal likelihood of the
constrained \MBA{} model to the marginal likelihood of the full \MBA{}
model. Then a Bayes factor greater than 1 would indicate that there is
no evidence in favour of the full \MBA{} model and therefore, the
background mark probabilities do not capture home advantage. However,
as a consequence of both \MBA{} models being high-dimensional
($\sim1500$ parameters), calculating their marginal likelihoods proved
computationally infeasible.

As an alternative, for the constrained \MBA{} model, we calculate its 
out-of-sample log predictive density on the same test data as carried 
out for all the other fitted models in Section~\ref{sec:modeval}. 
In fact, the constrained \MBA{} model ($\widehat{lpd} = -21589.28$) 
turns out with better predictive performance than the full \MBA{} 
model ($\widehat{lpd} = -21599.81$), supporting our claim that the 
background process of the game is not influenced by home advantage. 

\begin{figure}[!t]
\centering
\resizebox{0.9\linewidth}{!}{
\begin{subfigure}{0.4\textwidth}
\includegraphics[height=8.5cm, left]{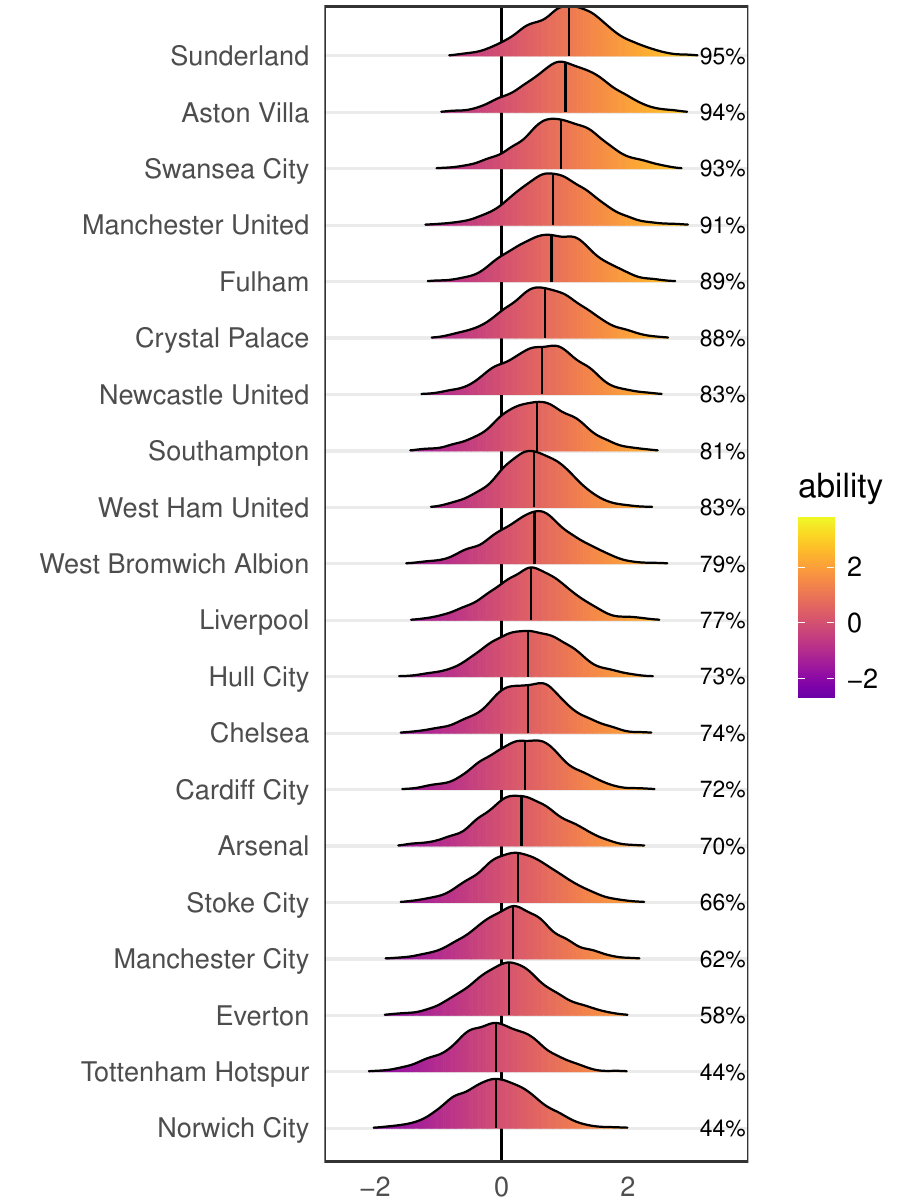} 
\caption{}
\label{figa:homeability}
\end{subfigure}
\begin{subfigure}{0.4\textwidth}
\includegraphics[height=8.5cm, right]{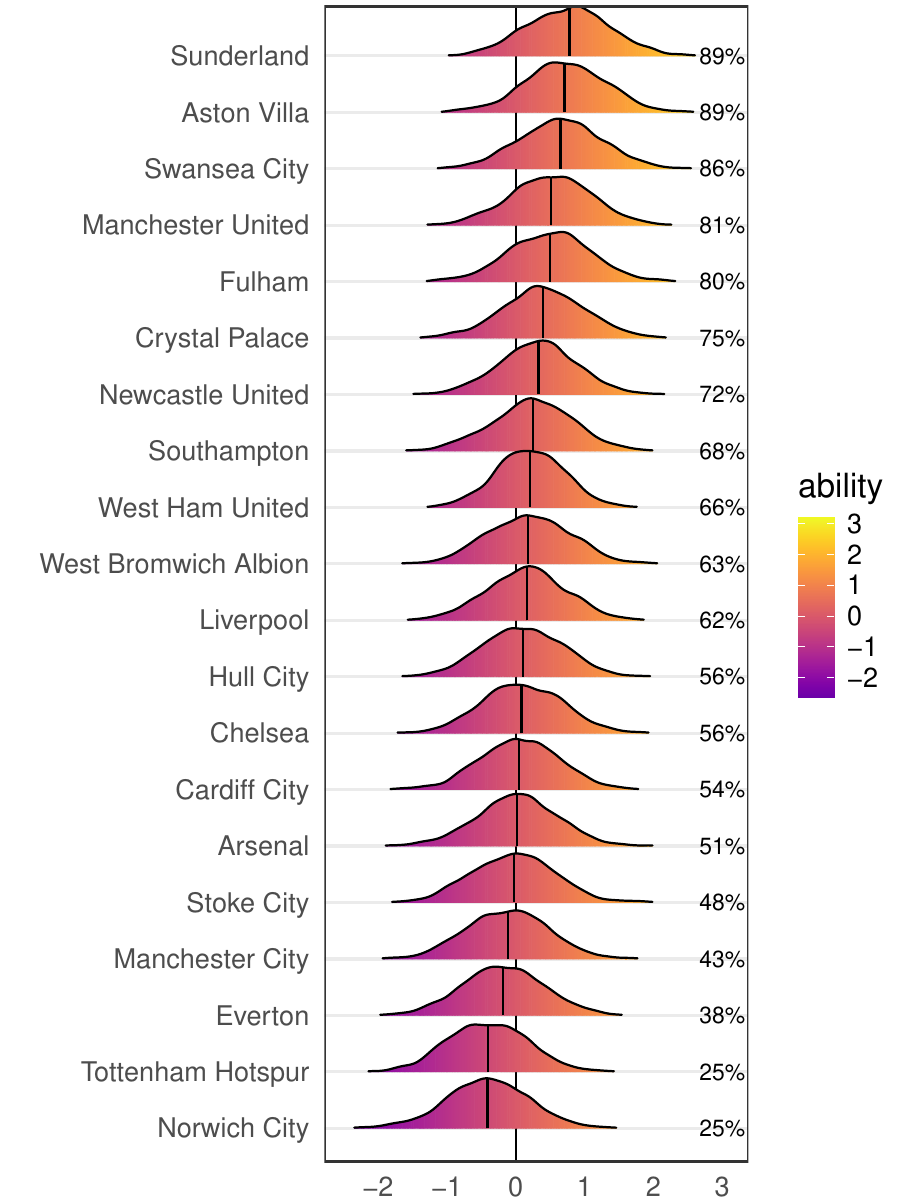}
\caption{}
\label{figb:homeability}
\end{subfigure}}
\caption{Posterior distribution of 
$\phi_{\textsf{Home\_Win} \to \textsf{Home\_Pass\_S} \mid 2}$ + $\omega_{c,\textsf{Home\_Pass\_S}}$ - $\phi_{\textsf{Away\_Win} \to \textsf{Away\_Pass\_S} \mid 2}$ - $\omega_{c,\textsf{Away\_Pass\_S}}$ in (a) 
and $\phi_{\textsf{Home\_Pass\_S} \to \textsf{Home\_Pass\_S} \mid 2}$ + $\omega_{c,\textsf{Home\_Pass\_S}}$ - $\phi_{\textsf{Away\_Pass\_S} \to \textsf{Away\_Pass\_S} \mid 2}$ - $\omega_{c,\textsf{Away\_Pass\_S}}$ 
in (b), from the baseline logit specification for incorporating team abilities 
in (\ref{baselogitspecteams}). Interpreted as the relative 
ability of a team to convert a \textsf{Win} to a \textsf{Successful Pass} 
when playing at home compared to away in (a) and similarly from one 
\textsf{Successful Pass} to another \textsf{Successful Pass} in (b). 
Teams are ranked in the  decreasing order of the means of their 
respective posterior distributions shown by the overlaid vertical lines.}
\label{fig:homeability}
\end{figure}

We also observe that the successful \textsf{Pass} events account for 
the majority of the background probability mass, while events like 
\textsf{Shots} and \textsf{Goals} have nearly zero probability. This 
suggests that the \textsf{Shot} and \textsf{Goal} events are highly 
unlikely to originate solely from the background component, but are 
instead triggered by excitations from previous events.

\subsection{Excitation factor}

The excitation factor $\alpha$ in expression~(\ref{modspecmarksmatrix}) 
is a scaling factor applied to the contributions from the previous 
occurrences to the event mark probability. In~(\ref{modspecmarksmatrix}), 
the background component has a weight of 1, while previous 
occurrences are weighted by $\exp(\alpha)$. 

The 95\% highest posterior density interval for $\exp(\alpha)$ is
$(451.35, 642.54)$, providing evidence that the contributions from
previous occurrences carry substantially higher weight relative to the
background component. In other words, this indicates that event
sequences in football have a significant dependence on their history.

\subsection{Decay rates}

As mentioned in Section~\ref{sec:modparinterpret} the decay rate
$\beta_{m \to m' \mid z}$ in expression (\ref{modspecmarksmatrix}) is
the exponential decay rate of the excitation caused by an event of
mark $m$ on an event of mark $m'$ at location $z$. By allowing the
decay rates to depend on the pair of marks involved in the excitation,
we had hoped to account for scenarios like a \textsf{Corner} event
exciting a \textsf{Pass\_S} event in the short term and a
\textsf{Shot} event in the longer term. Indeed, the 95\% highest 
posterior density interval for 
$\beta_{\textsf{Home\_Corner} \to \textsf{Home\_Pass\_S} \mid 3}$ 
is $(1.34, 2.36)$ and 
$\beta_{\textsf{Home\_Corner} \to \textsf{Home\_Shot} \mid 3}$ is 
$(0.16, 0.44)$ illustrating that the $\textsf{Corner} \to \textsf{Shot}$ 
excitation decays at a much slower rate compared to the 
$\textsf{Corner} \to \textsf{Pass\_S}$ excitation.

\begin{figure}[!t]
\centering
\resizebox{0.9\linewidth}{!}{
\begin{subfigure}{0.4\textwidth}
\includegraphics[height=8.5cm, left]{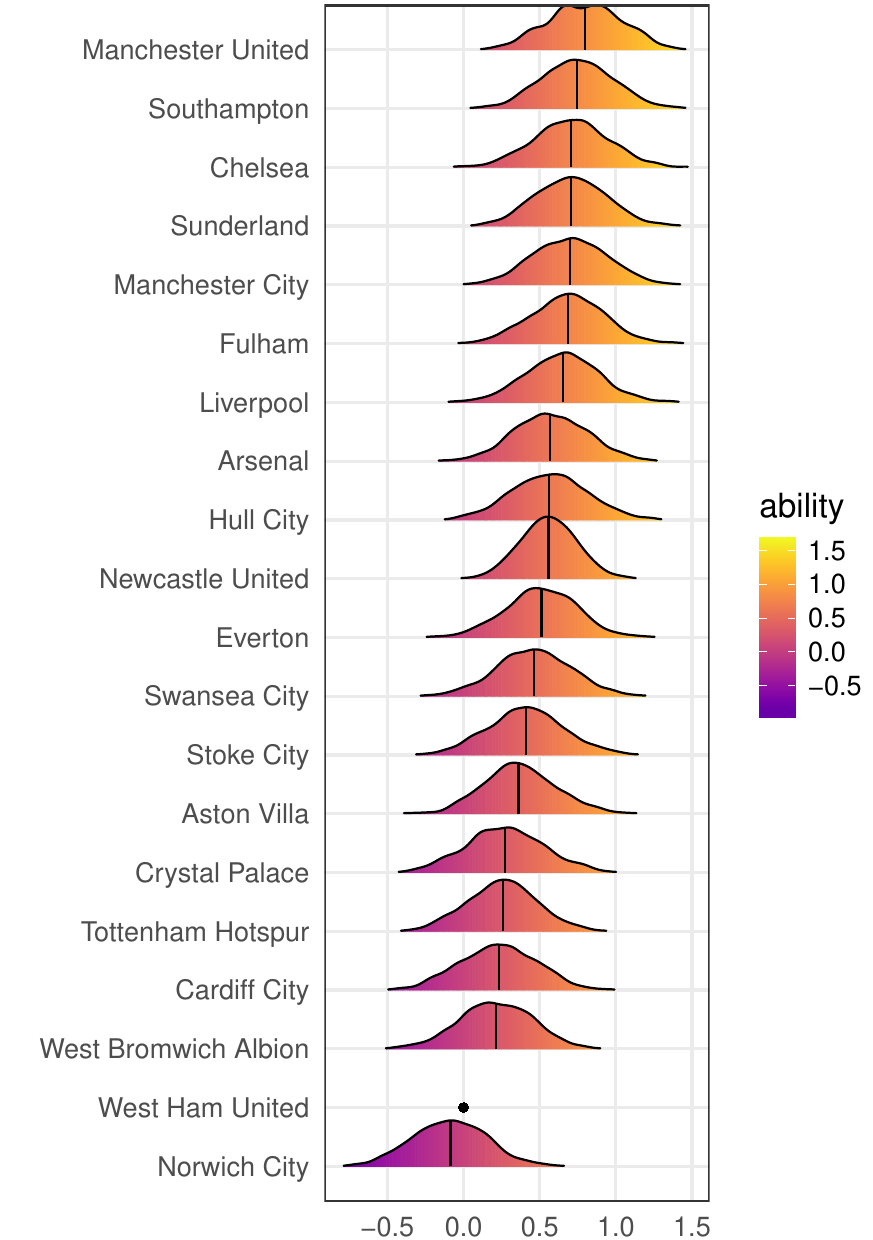} 
\caption{}
\label{figa:passability}
\end{subfigure}
\begin{subfigure}{0.4\textwidth}
\includegraphics[height=8.5cm, right]{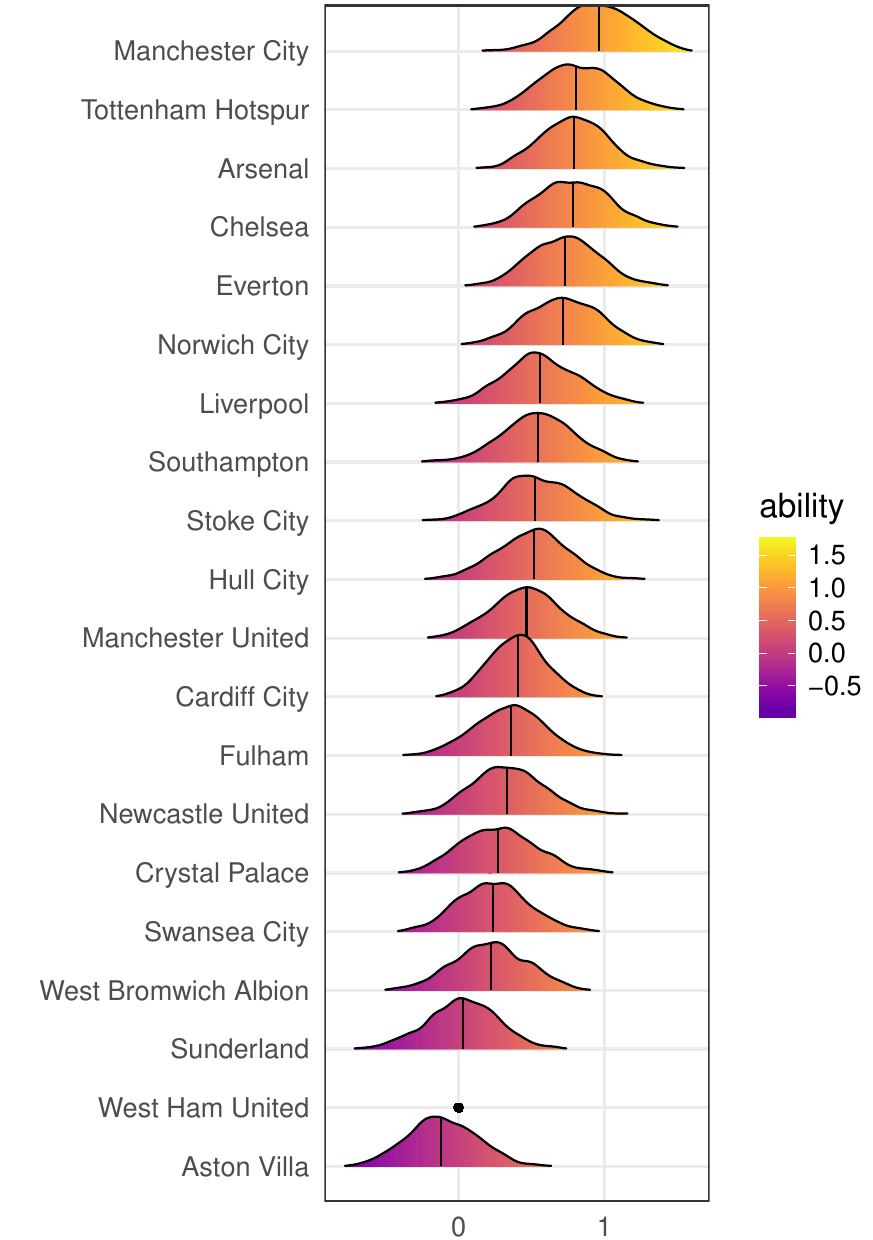}
\caption{}
\label{figb:passability}
\end{subfigure}}
\caption{Posterior distribution of the parameters $\omega_{c,\textsf{Home\_Pass\_S}}$ in (a) and $\omega_{c,\textsf{Away\_Pass\_S}}$ in (b), from the baseline logit specification for incorporating team abilities in expression (\ref{baselogitspecteams}). Teams are ranked in the decreasing order of the means of their respective posterior distributions shown by the overlaid vertical lines.}
\label{fig:passability}
\end{figure}

\subsection{Conversion rates}
\label{homeadvref}

The parameter $\gamma_{m \to m' \mid z}$ in
expression~(\ref{modspecmarksmatrix}) is the probability the
excitation from an event of mark $m$ triggers an event of mark $m'$ at
location $z$. 

Table~\ref{tab:gamma_zone} gives the posterior means and 
standard deviations of $\gamma_{m \to m' \mid z}$ in the midfield region 
($z = 2$) for Manchester United. The probabilities for the 
$\textsf{Home\_Win} \to \textsf{Home\_Pass\_S}$, 
$\textsf{Home\_Dribble} \to \textsf{Home\_Pass\_S}$ and
$\textsf{Home\_Pass\_S} \to \textsf{Home\_Pass\_S}$ conversions are
higher compared to their away team counterparts, indicating that 
Manchester United is better in retaining possession of the ball when 
playing at home compared to away. 

Figure~\ref{fig:homeability} provides a ridge-line plot of the log
odds ratio for home versus away ability of a team to convert a
$\textsf{ Win} \to \textsf{ Pass\_S}$ (Figure \ref{figa:homeability})
and $\textsf{Pass\_S} \to \textsf{ Pass\_S}$ (Figure
\ref{figb:homeability}).  The teams are listed in decreasing order of
the means of their respective posterior log odds ratios which are
indicated by vertical lines.  The percentage values by each plot,
indicate the fraction of the distribution greater than 0. All but two
teams in (Figure \ref{figa:homeability}) and five teams in (Figure
\ref{figb:homeability}) have greater than $50\%$ of their distribution
greater than 0, confirming that the vast majority of teams possess a
higher ability to retain possession while playing at home.

In this way, we not only confirm the well-known home advantage effect, 
but also quantify team performance for games played at home as well
as away.

\subsection{Team abilities}

Figure~\ref{fig:passability} provides a ridge-line plot of the
posterior distribution of the parameters
$\omega_{c,\textsf{Home\_Pass\_S}}$ and
$\omega_{c,\textsf{Away\_Pass\_S}}$. The teams are listed in the
decreasing order of the means of their respective posterior
distributions which are indicated by vertical lines. We observe that
Manchester United, the team with the highest ability to retain
possession in home games (Figure \ref{figa:passability}), drop
significantly down in the rankings for the away games (Figure
\ref{figb:passability}). This is evidence that when Manchester 
United plays away they seem to deviate from the possession-based 
strategy they seem to adopt in the home games.

\begin{figure}[!t]
\centering
\resizebox{0.9\linewidth}{!}{
\begin{subfigure}{0.4\textwidth}
\hspace{-4mm}
\includegraphics[height=8.5cm, left]{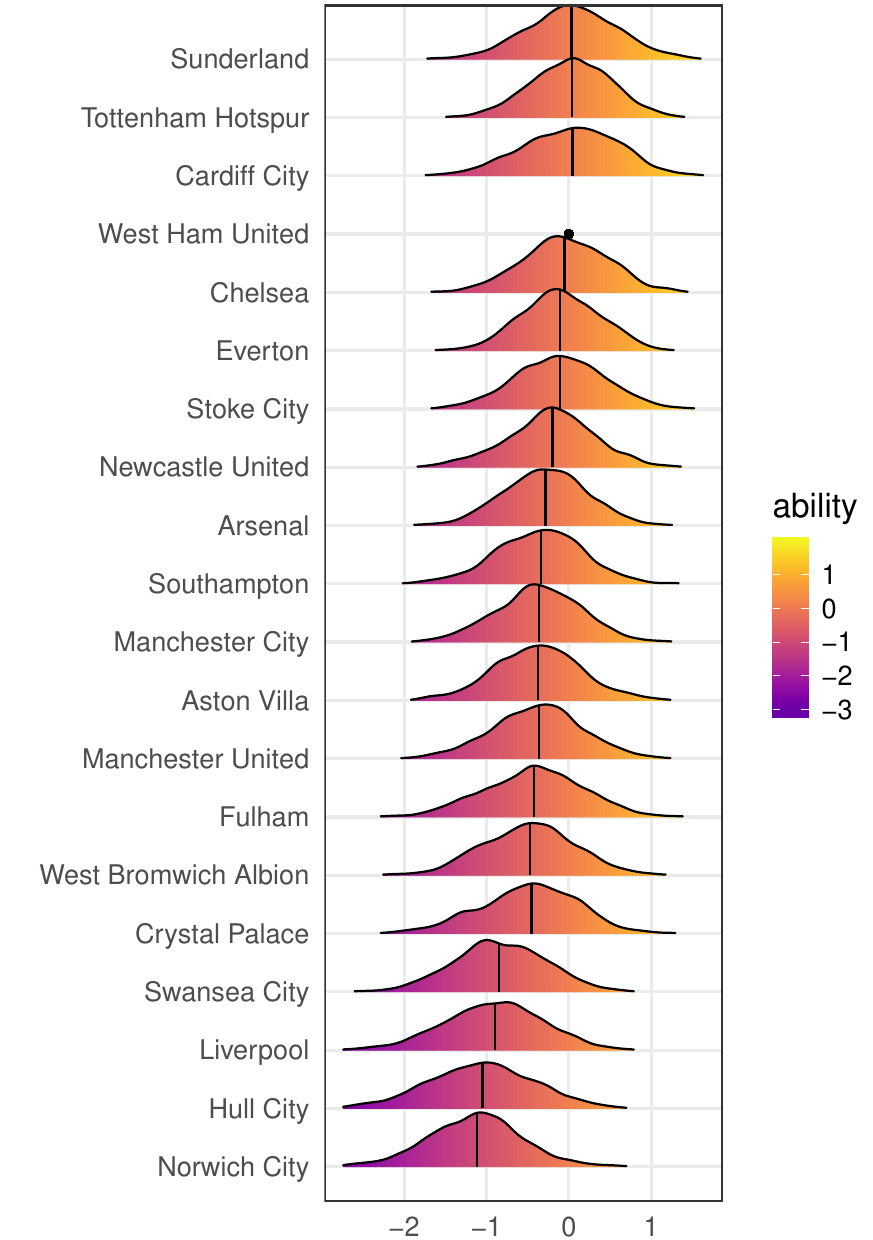} 
\caption{}
\label{figa:shotability}
\par\vspace{0pt}
\end{subfigure}
\hfill
\begin{subfigure}{0.55\textwidth}
\footnotesize
\raggedleft
\hspace{-2mm}
\begin{tabular}[t]{lccc}
\toprule
Team & Shots & Passes & S/P\\
\midrule
Cardiff City & 18 & 115 & 0.16\\
Sunderland & 28 & 195 & 0.14\\
Tottenham Hotspur & 37 & 261 & 0.14\\
Aston Villa & 20 & 153 & 0.13\\
Newcastle United & 22 & 180 & 0.12\\
Crystal Palace & 18 & 148 & 0.12\\
Stoke City & 24 & 199 & 0.12\\
Everton & 40 & 342 & 0.12\\
Chelsea & 29 & 248 & 0.12\\
Fulham & 19 & 163 & 0.12\\
West Ham United & 22 & 194 & 0.11\\
West Bromwich Albion & 18 & 159 & 0.11\\
Swansea City & 23 & 214 & 0.11\\
Arsenal & 30 & 280 & 0.11\\
Southampton & 23 & 223 & 0.10\\
Liverpool & 28 & 292 & 0.10\\
Manchester United & 22 & 238 & 0.09\\
Manchester City & 30 & 330 & 0.09\\
Norwich City & 19 & 227 & 0.08\\
Hull City & 13 & 191 & 0.07\\
\bottomrule
\end{tabular}
    \par\vspace{0pt}
\caption{}
\label{figb:shotability}
\end{subfigure}}
\caption{(a) Posterior distribution of $\omega_{c,\textsf{Home\_Shot}} + \omega_{c,\textsf{Away\_Shot}}$, the cumulative ability of a team $c$, relative to West Ham (baseline), to attempt a shot on goal. (b) The number of shots, passes completed in the attacking third and shots per pass completed in the attacking third (S/P) for each team in the training data.}
\label{fig:shotability}
\end{figure}

Figure \ref{figa:shotability} provides a ridge-line plot of the
posterior distribution of the cumulative ability of a team to attempt
a shot on goal. A higher $\omega_{c,\textsf{Home\_Shot}}$, for
example, indicates that for the team $c$, an event like
$\textsf{Home\_Pass\_S}$ is more likely to trigger a
$\textsf{Home\_Shot}$. We do not expect the cumulative abilities
$\omega_{c,\textsf{Home\_Shot}} + \omega_{c,\textsf{Away\_Shot}}$ of
the dominant teams to be high, as they might prefer to make additional
passes to create better goal scoring opportunities. A weaker team, on
the other hand, typically has fewer opportunities to attack and
therefore, is more likely to attempt a shot on goal when
possible. Indeed, this is what we observe in
Figure~\ref{fig:shotability}, where we compare the team rankings based
on their cumulative ability
$\omega_{c,\textsf{Home\_Shot}} + \omega_{c,\textsf{Away\_Shot}}$ with
the number of shots per pass completed in the attacking third (S/P
column in Figure \ref{figb:shotability}) in the training data. The
comparison between Cardiff City and Norwich City is an interesting
example of two teams that appear to be similar with 18 and 19 shots on
goal attempted, respectively, in their two games in the training
data. However, the two teams are at the opposite ends of the ranking
based on their cumulative ability
$\omega_{c,\textsf{Home\_Shot}} + \omega_{c,\textsf{Away\_Shot}}$,
capturing the clear difference between their attacking styles.

\begin{table}[!t]
\centering
\caption{(a) Team rankings based on the cumulative ability to trigger a particular event type. For example, the column \textsf{Pass} ranks teams in the decreasing order of their posterior means of $\omega_{c,\textsf{Home\_Pass\_S}} + \omega_{c,\textsf{Away\_Pass\_S}}$. The teams are ordered in (a) by the rankings based on their passing ability, which is a good indicator of the final position in the league table of the 2013/14 season in (b).}
\resizebox{0.9\linewidth}{!}{
    \begin{subtable}{.6\linewidth}
    \caption{}
    \footnotesize
    \raggedright
    \textsf{
    \begin{tabular}[t]{lccccc}
    \toprule
    Team & Pass & Shot & Goal & Win & Save\\
    \midrule
    Manchester City & 1 & 11 & 1 & 15 & 11\\
    Chelsea & 2 & 5 & 11 & 20 & 4\\
    Arsenal & 3 & 9 & 3 & 5 & 7\\
    Southampton & 4 & 10 & 8 & 7 & 19\\
    Manchester United & 5 & 13 & 4 & 2 & 18\\
    Everton & 6 & 6 & 17 & 11 & 14\\
    Liverpool & 7 & 18 & 12 & 9 & 5\\
    Hull City & 8 & 19 & 15 & 1 & 10\\
    Tottenham Hotspur & 9 & 2 & 14 & 3 & 6\\
    Fulham & 10 & 14 & 7 & 14 & 3\\
    Stoke City & 11 & 7 & 9 & 12 & 8\\
    Newcastle United & 12 & 8 & 19 & 16 & 17\\
    Sunderland & 13 & 1 & 13 & 8 & 2\\
    Swansea City & 14 & 17 & 18 & 17 & 12\\
    Cardiff City & 15 & 3 & 2 & 19 & 15\\
    Norwich City & 16 & 20 & 10 & 4 & 20\\
    Crystal Palace & 17 & 16 & 16 & 13 & 16\\
    West Bromwich Albion & 18 & 15 & 20 & 10 & 1\\
    Aston Villa & 19 & 12 & 5 & 6 & 13\\
    West Ham United & 20 & 4 & 6 & 18 & 9\\
    \bottomrule
    \end{tabular}}
  \par\vspace{0pt}
  \label{taba:rankings}
    \end{subtable}%
    \hfill
    \begin{subtable}{.4\linewidth}
    \caption{}
        \footnotesize
  \raggedleft
  \textsf{
    \begin{tabular}[t]{cl}
    \toprule
    League Position & Team\\
    \midrule
    1 & Manchester City\\
    2 & Liverpool\\
    3 & Chelsea\\
    4 & Arsenal\\
    5 & Everton\\
    6 & Tottenham Hotspur\\
    7 & Manchester United\\
    8 & Southampton\\
    9 & Stoke City\\
    10 & Newcastle United\\
    11 & Crystal Palace\\
    12 & Swansea City\\
    13 & West Ham United\\
    14 & Sunderland\\
    15 & Aston Villa\\
    16 & Hull City\\
    17 & West Bromwich Albion\\
    18 & Norwich City\\
    19 & Fulham\\
    20 & Cardiff City\\
    \bottomrule
    \end{tabular}}
  \par\vspace{0pt}
  \label{tabb:rankings}
    \end{subtable}}
    \label{tab:rankings}
\end{table}

Table~\ref{taba:rankings} shows the team rankings based on the
cumulative ability to trigger five different event types. For example,
the \textsf{Pass} column ranks teams in the decreasing order of their
posterior means of
$\omega_{c,\textsf{Home\_Pass\_S}} +
\omega_{c,\textsf{Away\_Pass\_S}}$. The teams are ordered in
Table~\ref{taba:rankings} by the rankings based on their cumulative
passing ability. Despite training on just the first 20 out of 380
games of the 2013/14 season, the rankings based on the passing ability
is a good indicator of the positions the teams finished in the final
league table of the 2013/14 season in Table~\ref{tabb:rankings}.

\subsection{Event genealogy}
\label{sec::eventgene}

\begin{figure}[!t]
\centering
\resizebox{\linewidth}{!}{
\includegraphics {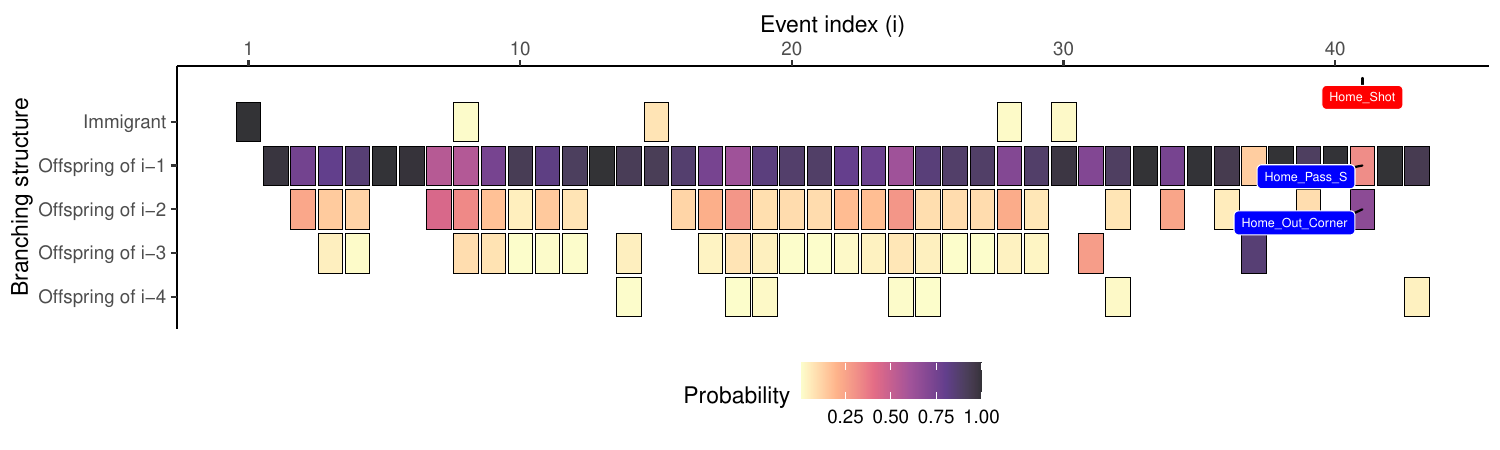}}
\caption{Posterior means of branching structure probabilities for events in the first 4 minutes of the game between Chelsea and Hull City on 18/08/2013. The highlighted event \textsf{Home\_Shot} has a higher probability of being an offspring of the event \textsf{Home\_Out\_Corner} than being an offspring of the more recent \textsf{Home\_Pass\_S} event.}
\label{fig:branchingstruc}
\end{figure}

The branching structure $u_{si}$ indicates whether the $i$th event in $s$th 
sequence is an ``immigrant'' $(u_{si} = 0)$ or an ``offspring'' of a previous 
event with index $j$ $(u_{si} = j)$. Given an observed event sequence
$\mathcal{F}_{st_{sn_s}}$, the conditional branching structure probabilities
$\mathbb{P}(u_{si} \mid \mathcal{F}_{st_{si}})$ based on the model
specification in expression (\ref{modspecmarksmatrix}) are
\begin{align}
\mathbb{P}(u_{si} = 0 \mid \mathcal{F}_{st_{si}}) &= \frac{\delta_{m_{si} \mid z_{si}}}{\delta_{m_{si} \mid z_{si}} + \sum_{t_{sk} < t_{si}} \mathrm{e}^{\alpha - \beta_{m_{sk} \to m_{si} \mid z_{si}} (t_{si}-t_{sk})}\gamma_{m_{sk} \to m_{si} \mid z_{si}}} \, , \nonumber \\
\mathbb{P}(u_{si} = j \mid \mathcal{F}_{st_{si}}) &= 
\begin{cases}
\frac{\mathrm{e}^{\alpha - \beta_{m_{sj} \to m_{si} \mid z_{si} } (t_{si}-t_{sj})}\gamma_{m_{sj} \to m_{si} \mid z_{si} }}{\delta_{m_{si}} + \sum_{t_{sk} < t_{si}} \mathrm{e}^{\alpha - \beta_{m_{sk} \to m_{si} \mid z_{si}} (t_{si}-t_{sk})} \gamma_{m_{sk} \to m_{si} \mid z_{si}}}& \text{for $t_{sj} < t_{si}$}\\
0& \text{for $t_{sj} \geq t_{si}$}
\end{cases}  \, .
\label{exp:branch_struc_probs_matrix}
\end{align}

The branching structure probabilities
in~(\ref{exp:branch_struc_probs_matrix}) quantify the relative
contributions of the background process and previous occurrences in
the mark probability of the $i$th event in the $s$th sequence. Figure
\ref{fig:branchingstruc} shows the posterior means of the branching
structure probabilities for all events in the first four minutes of
the game between Chelsea and Hull City on 18/08/2013. To illustrate
the flexibility of the model to account for dependence between events
over arbitrary durations of time, we highlight the event
\textsf{Home\_Shot} showing a higher probability of being an offspring
of the event \textsf{Home\_Out\_Corner} than being an offspring of the
more recent \textsf{Home\_Pass\_S} event.

\section{Model-based predictions}

Finally, we illustrate how the mechanistic modelling framework presented 
in this paper can be used to simulate event sequences in football and 
obtain predictions of event probabilities in real-time. We split the game 
between Arsenal and Tottenham Hotspur (01/09/2013) in the 
test data into 30-second intervals. For each interval, given the history of 
events up to but not including the interval, we simulate events over the 
next 30 seconds $Q = 100$ times for each of the $R = 500$ posterior 
samples from the \MBA{} model with the tuning parameter setting 
($W = 5$, $N = 100$).

\begin{figure}[!t]
\centering
\resizebox{0.95\linewidth}{!}{
\includegraphics{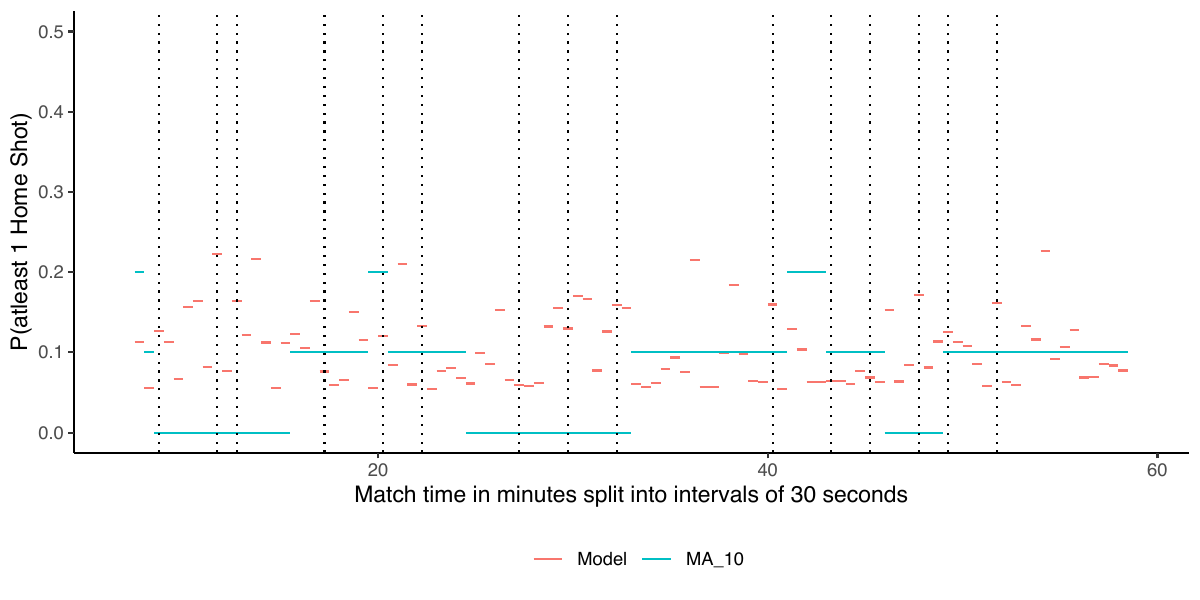}}
\caption{Forecasting the probability of observing at least one \textsf{Home 
Shot} event in 30-second intervals during the game between 
Arsenal and Tottenham Hotspur (01/09/2013) in the test data. Intervals 
with observed \textsf{Home Shot} events are highlighted using dotted 
lines. \textsf{MA\_10} is a 10-step moving average model used as a 
benchmark for comparison.}
\label{fig:shotevo}
\end{figure}

In Figure \ref{fig:shotevo}, we plot the proportion of all simulations
within each interval where at least one \textsf{Home\_Shot} event was
simulated, and use dotted lines to denote the intervals where a
\textsf{Home\_Shot} event was actually observed. We also include a
10-step moving average model (\textsf{MA\_10}) as a benchmark for
comparison. We excluded the first 5 minutes of the game to ensure that
we have predictions from both the models being compared. A quick
inspection reveals that in 11 of the 15 intervals in which a
\textsf{Home\_Shot} is observed, the model predicts a shot probability
greater than the 10-step moving average model.

\begin{figure}[!t]
\centering
\resizebox{0.9\linewidth}{!}{
\includegraphics{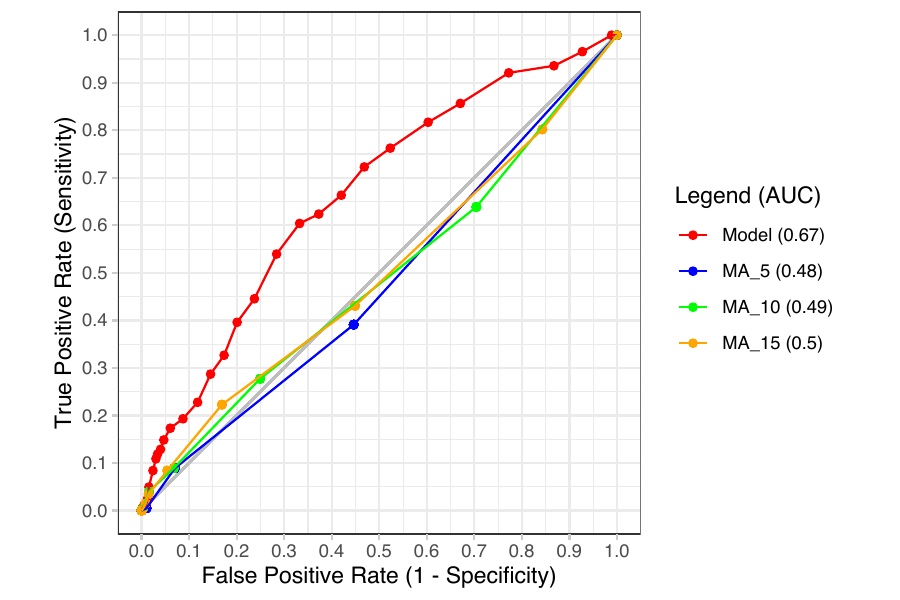}}
\caption{Validating the model performance against three moving 
average models for the task of whether a shot will be observed in each 
30-second interval over the first 20 games in the test set. 
\textsf{MA\_5} is a 5-step moving average model and so on. The ROC curve 
evaluates the performance of a classification model over all classification 
thresholds. The area under the curve (AUC) values in the legend clearly 
confirm the superior performance of the model.}
\label{fig:shotroc}
\end{figure}

In Figure \ref{fig:shotroc}, we formally validate the performance of the 
model against three moving average models for the classification task of 
whether a shot will be observed in an interval. For this purpose, we use 
data from the first 20 games in the test set, where we excluded the first 15 
intervals of each game to ensure that we have predictions from all the four 
models being compared. To validate the models, we had a total of 1959 
intervals out of which 202 intervals had at least one \textsf{Home\_Shot} 
event. The area under the Receiver Operating Characteristic (ROC) curve 
is a performance measure that evaluates the performance of a classification 
model over all classification thresholds. The area under the curve (AUC) values 
are given in the legend and clearly confirm the superior performance of the 
model.

\section{Discussion and concluding remarks}

Building on the decomposition of a multivariate distribution function, we 
showed how the joint modelling in classical point process models like 
Hawkes processes, can be decoupled. The introduced flexible modelling 
framework can, for example, retain the characteristic property of 
excitation in Hawkes processes in the model for the marks while avoiding 
the clustering of event times. A comprehensive Bayesian approach for the 
modelling of flexible marked spatio-temporal point processes was 
developed including an approach to evaluate the predictive accuracy of 
the fitted Bayesian models using the out-of-sample log predictive density.

We presented a case study showing how the modelling framework
developed in this paper can be tailored to separately model the
components of the events in football, namely, the times, the locations
and the event types. We were also able to incorporate team information
into the model in a direct way that captured the relative abilities of
the teams for each event type. We developed a method based on
association rules to reduce the increased model complexity introduced
by model extensions. The rule-based approach identified significant
event interactions within sequences by placing thresholds on measures
of significance. We then evaluated the accuracy of the excitation
based models by comparing against two baseline models and confirmed
the superior performance of the models with excitation effects.

We provided a detailed parameter description showing how the model
parameters can be used to gain valuable insight into football. The
excitation framework of the best performing model captured both the
magnitudes and the durations of all pairwise event interactions across
different locations. From the conversion rate parameters, we were able
to quantify the well-known home advantage effect. We also discussed
how the team ability parameters can be used to obtain rankings for the
teams by event type, that can be used as predictors for team
performance. The team ability parameters also captured some
interesting differences in the playing styles of the teams, that were
not immediately apparent just by looking at the data. In this way, the
model along with its parameters can be used to develop a deeper
understanding of the game-play by the coaching staff and inform
strategic decision making. The proposed model can also be used to
simulate the sequence of events in a game to obtain real-time
predictions of event probabilities. We believe these predictions would
enhance, among other aspects, the viewing experience of televised
games.

The dataset we used consists of events from a single English Premier
League season, which has a total of 380 games. However, as described
in Section~\ref{sec:modtrain}, we only used the first 20 games of the
season as training data for the modelling exercise, over which every
team in the league plays exactly one game each at Home and Away
venues. This represents the minimum number of games required to ensure
identifiability of all model parameters, specifically the team
abilities. Even though the volume of data was kept to the minimum for
computational reasons, our results illustrate that the model can
provide valuable insights with limited data. So, the methodology
developed in this paper can be readily applied to other team sports
like rugby, hockey, basketball, American football etc where there may
be fewer events per game or fewer games in a season.

Multiple seasons can be modelled together as if it were just one
season using the modelling framework we propose, as long as the game
rules, and hence, the definition of the events being considered does
not change. Another aspect of the tournament to note is the relegation
and promotion of teams within the league, which results in some teams
not playing the same number of games over multiple seasons. A
limitation of the proposed model is that the game periods are
exchangeable, because the likelihood is invariant to the order in
which the game periods and the games occur. It would be more natural
to allow for the team ability parameters
in~(\ref{exp:baselogitspecbayes}) to be time-varying, especially over
multiple seasons during which team players and managers are likely to
change. Due to computational reasons we were not able to utilise most
of the data even within a single season and current work focuses on
overcoming this computational barrier using variational inference
\citep{blei:2017}.
 
As none of the methods have been tailored specifically to football or
even sports for that matter, they can be applied to a wide range of
applications that generate event data streams. Specifically, the
conversion rate parameters can be used to capture the triggering
structure between different event types, for example, the probability
of large earthquakes triggering smaller aftershocks. Also, the team
ability parameters can be used for other multi-agent environments, for
example, accidents by car type, countries in the analyses of financial
events, individuals in identity systems, and so on.

\section{Supporting materials}
The raw data that motivated this work has been provided by Stratagem
Technologies Ltd and consists of all touch-ball events for the 2013/14
season of the English Premier League. The raw data cannot be disclosed
as the authors do not have the license to do so. The GitHub repository
\url{https://github.com/ForeStats/flexible-msttp-football}  provides all the
computer code used for the data pre-processing and the \textsf{Stan}
templates used to carry out the analyses presented in this paper. We
also provide guidance on how the code can be used to apply our methods
to similar data sets, like the publicly-available 2020/21 FA Women's
Super League Data provided by StatsBomb Inc. at
\url{https://github.com/statsbomb/open-data}. Section~S1 of the
Supporting Materials document provides details on the prior
distributions and the derivation of their posterior distributions for
the \FOMC{} and \MSTHP{} models. Section~S2 gives details on the
association rule based screening procedure used to identify the most
significant event interactions. We also provide the chain-wise trace
plots for some of the parameters of the \MBA{} model with $W = 5$ and
$N = 100$.

\section{Acknowledgements}
The authors thank Stratagem Technologies Ltd for giving access to all
touch-ball events from all English Premier League games in the 2013/14
season.

\bibliographystyle{chicago}
\setcitestyle{authoryear,open={(},close={)}}
\bibliography{ref}

\begin{thebibliography}{}

\bibitem[\protect\citeauthoryear{Agrawal, Imieliundefinedski, and
  Swami}{Agrawal et~al.}{1993}]{agarwal1993mining}
Agrawal, R., T.~Imieliundefinedski, and A.~Swami (1993).
\newblock Mining association rules between sets of items in large databases.
\newblock {\em SIGMOD Rec.\/}~{\em 22\/}(2), 207--216.

\bibitem[\protect\citeauthoryear{Agresti}{Agresti}{2007}]{agresti2007categorical}
Agresti, A. (2007).
\newblock {\em An Introduction to Categorical Data Analysis\/} (2nd ed.).
\newblock Hoboken, NJ: Wiley.

\bibitem[\protect\citeauthoryear{Blei, Kucukelbir, and McAuliffe}{Blei
  et~al.}{2017}]{blei:2017}
Blei, D.~M., A.~Kucukelbir, and J.~D. McAuliffe (2017).
\newblock Variational inference: A review for statisticians.
\newblock {\em Journal of the American statistical Association\/}~{\em
  112\/}(518), 859--877.

\bibitem[\protect\citeauthoryear{Borrie, Jonsson, and Magnusson}{Borrie
  et~al.}{2002}]{borrie2002temporal}
Borrie, A., G.~K. Jonsson, and M.~S. Magnusson (2002).
\newblock Temporal pattern analysis and its applicability in sport: an
  explanation and exemplar data.
\newblock {\em Journal of sports sciences\/}~{\em 20\/}(10), 845--852.

\bibitem[\protect\citeauthoryear{Bowsher}{Bowsher}{2007}]{bowsher2007modelling}
Bowsher, C.~G. (2007).
\newblock Modelling security market events in continuous time: Intensity based,
  multivariate point process models.
\newblock {\em Journal of Econometrics\/}~{\em 141\/}(2), 876--912.

\bibitem[\protect\citeauthoryear{Clemente, Couceiro, Martins, and
  Mendes}{Clemente et~al.}{2015}]{clemente2015using}
Clemente, F.~M., M.~S. Couceiro, F.~M.~L. Martins, and R.~S. Mendes (2015).
\newblock Using network metrics in soccer: a macro-analysis.
\newblock {\em Journal of human kinetics\/}~{\em 45\/}(1), 123--134.

\bibitem[\protect\citeauthoryear{Cox}{Cox}{1975}]{cox1975partial}
Cox, D.~R. (1975).
\newblock Partial likelihood.
\newblock {\em Biometrika\/}~{\em 62\/}(2), 269--276.

\bibitem[\protect\citeauthoryear{Daley and Vere-Jones}{Daley and
  Vere-Jones}{2003}]{daley2003theory}
Daley, D.~J. and D.~Vere-Jones (2003).
\newblock {\em An introduction to the theory of point processes. {V}ol. {I}\/}
  (2nd ed.).
\newblock New York: Springer-Verlag.

\bibitem[\protect\citeauthoryear{{Decroos}, {Bransen}, {Van Haaren}, and
  {Davis}}{{Decroos} et~al.}{2018}]{decroos2018actions}
{Decroos}, T., L.~{Bransen}, J.~{Van Haaren}, and J.~{Davis} (2018).
\newblock {Actions Speak Louder Than Goals: Valuing Player Actions in Soccer}.
\newblock arXiv:1802.07127.

\bibitem[\protect\citeauthoryear{Decroos, Dzyuba, Haaren, and Davis}{Decroos
  et~al.}{2017}]{decroos2017predicting}
Decroos, T., V.~Dzyuba, J.~V. Haaren, and J.~Davis (2017).
\newblock Predicting soccer highlights from spatio-temporal match event
  streams.
\newblock In S.~P. Singh and S.~Markovitch (Eds.), {\em AAAI}, pp.\
  1302--1308.

\bibitem[\protect\citeauthoryear{Diggle}{Diggle}{1985}]{diggle1985kernel}
Diggle, P. (1985).
\newblock A kernel method for smoothing point process data.
\newblock {\em Journal of the Royal Statistical Society: Series C (Applied
  Statistics)\/}~{\em 34\/}(2), 138--147.

\bibitem[\protect\citeauthoryear{Diggle}{Diggle}{2013}]{diggle2013spatial}
Diggle, P.~J. (2013).
\newblock {\em Statistical Analysis of Spatial and Spatio-Temporal Point
  Patterns\/} (3rd ed.).
\newblock Boca Raton, Florida: CRC Press.

\bibitem[\protect\citeauthoryear{Duane, Kennedy, Pendleton, and Roweth}{Duane
  et~al.}{1987}]{duane1987hybrid}
Duane, S., A.~D. Kennedy, B.~J. Pendleton, and D.~Roweth (1987).
\newblock Hybrid monte carlo.
\newblock {\em Physics letters B\/}~{\em 195\/}(2), 216--222.

\bibitem[\protect\citeauthoryear{Duch, Waitzman, and Amaral}{Duch
  et~al.}{2010}]{duch2010quantifying}
Duch, J., J.~S. Waitzman, and L.~A.~N. Amaral (2010).
\newblock Quantifying the performance of individual players in a team activity.
\newblock {\em PLOS One\/}~{\em 5\/}(6), e10937+.

\bibitem[\protect\citeauthoryear{Gelman, Carlin, Stern, Dunson, Vehtari, and
  Rubin}{Gelman et~al.}{2013}]{gelman2013bayesian}
Gelman, A., J.~Carlin, H.~Stern, D.~Dunson, A.~Vehtari, and D.~Rubin (2013).
\newblock {\em Bayesian Data Analysis\/} (3rd ed.).
\newblock Boca Raton, Florida: CRC Press.

\bibitem[\protect\citeauthoryear{Gelman, Rubin, et~al.}{Gelman
  et~al.}{1992}]{gelman1992inference}
Gelman, A., D.~B. Rubin, et~al. (1992).
\newblock Inference from iterative simulation using multiple sequences.
\newblock {\em Statistical science\/}~{\em 7\/}(4), 457--472.

\bibitem[\protect\citeauthoryear{Gonz{\'a}lez, Rodr{\'\i}guez-Cort{\'e}s,
  Cronie, and Mateu}{Gonz{\'a}lez et~al.}{2016}]{gonzalez2016spatio}
Gonz{\'a}lez, J.~A., F.~J. Rodr{\'\i}guez-Cort{\'e}s, O.~Cronie, and J.~Mateu
  (2016).
\newblock Spatio-temporal point process statistics: a review.
\newblock {\em Spatial Statistics\/}~{\em 18}, 505--544.

\bibitem[\protect\citeauthoryear{Grund}{Grund}{2012}]{grund2012network}
Grund, T.~U. (2012).
\newblock Network structure and team performance: The case of english premier
  league soccer teams.
\newblock {\em Social Networks\/}~{\em 34\/}(4), 682--690.

\bibitem[\protect\citeauthoryear{Gudmundsson and Horton}{Gudmundsson and
  Horton}{2017}]{gudmundsson2017spatio}
Gudmundsson, J. and M.~Horton (2017).
\newblock Spatio-temporal analysis of team sports.
\newblock {\em ACM Computing Surveys (CSUR)\/}~{\em 50\/}(2), 1--21.

\bibitem[\protect\citeauthoryear{Hawkes}{Hawkes}{1971}]{hawkes1971spectra}
Hawkes, A.~G. (1971).
\newblock Spectra of some self-exciting and mutually exciting point processes.
\newblock {\em Biometrika\/}~{\em 58\/}(1), 83--90.

\bibitem[\protect\citeauthoryear{Hawkes and Oakes}{Hawkes and
  Oakes}{1974}]{hawkes1974cluster}
Hawkes, A.~G. and D.~Oakes (1974).
\newblock A cluster process representation of a self-exciting process.
\newblock {\em Journal of Applied Probability\/}~{\em 11\/}(03), 493--503.

\bibitem[\protect\citeauthoryear{Hoffman and Gelman}{Hoffman and
  Gelman}{2014}]{hoffman2014no}
Hoffman, M.~D. and A.~Gelman (2014).
\newblock The no-u-turn sampler: Adaptively setting path lengths in hamiltonian
  monte carlo.
\newblock {\em Journal of Machine Learning Research\/}~{\em 15\/}(1),
  1593--1623.

\bibitem[\protect\citeauthoryear{Lindqvist}{Lindqvist}{2006}]{lindqvist:2006}
Lindqvist, B.~H. (2006).
\newblock On the statistical modeling and analysis of repairable systems.
\newblock {\em Statistical Science\/}~{\em 21\/}(4), 532--551.

\bibitem[\protect\citeauthoryear{Mackay}{Mackay}{2017}]{mackay2017predicting}
Mackay, N. (2017).
\newblock Predicting goal probabilities for possessions in football.
\newblock Master's thesis, Vrije Universiteit Amsterdam.

\bibitem[\protect\citeauthoryear{Mohler, Short, Brantingham, Schoenberg, and
  Tita}{Mohler et~al.}{2011}]{mohler2011self}
Mohler, G.~O., M.~B. Short, P.~J. Brantingham, F.~P. Schoenberg, and G.~E. Tita
  (2011).
\newblock Self-exciting point process modeling of crime.
\newblock {\em Journal of the American Statistical Association\/}~{\em
  106\/}(493), 100--108.

\bibitem[\protect\citeauthoryear{Narayanan}{Narayanan}{2021}]{narayanan:2021}
Narayanan, S. (2021).
\newblock {\em Bayesian Modelling of Flexible Marked Point Processes with
  Applications to Event Sequences from Association Football}.
\newblock Ph.\ D. thesis, University of Warwick, Coventry.
\newblock \url{http://webcat.warwick.ac.uk/record=b3520069\~S15}.

\bibitem[\protect\citeauthoryear{Norris}{Norris}{1997}]{norris1997markov}
Norris, J.~R. (1997).
\newblock {\em Markov Chains}.
\newblock Cambridge: Cambridge University Press.

\bibitem[\protect\citeauthoryear{Ogata}{Ogata}{1998}]{ogata1998space}
Ogata, Y. (1998).
\newblock Space-time point-process models for earthquake occurrences.
\newblock {\em Annals of the Institute of Statistical Mathematics\/}~{\em
  50\/}(2), 379--402.

\bibitem[\protect\citeauthoryear{Passos, Davids, Ara{\'u}jo, Paz, Mingu{\'e}ns,
  and Mendes}{Passos et~al.}{2011}]{passos2011networks}
Passos, P., K.~Davids, D.~Ara{\'u}jo, N.~Paz, J.~Mingu{\'e}ns, and J.~Mendes
  (2011).
\newblock Networks as a novel tool for studying team ball sports as complex
  social systems.
\newblock {\em Journal of Science and Medicine in Sport\/}~{\em 14\/}(2),
  170--176.

\bibitem[\protect\citeauthoryear{Pena and Touchette}{Pena and
  Touchette}{2012}]{pena2012network}
Pena, J.~L. and H.~Touchette (2012).
\newblock A network theory analysis of football strategies.
\newblock arXiv:1206.6904.

\bibitem[\protect\citeauthoryear{Rasmussen}{Rasmussen}{2013}]{rasmussen2013bayesian}
Rasmussen, J.~G. (2013).
\newblock Bayesian inference for hawkes processes.
\newblock {\em Methodology and Computing in Applied Probability\/}~{\em
  15\/}(3), 623--642.

\bibitem[\protect\citeauthoryear{Ripley}{Ripley}{1977}]{ripley1977modelling}
Ripley, B.~D. (1977).
\newblock Modelling spatial patterns.
\newblock {\em Journal of the Royal Statistical Society: Series B
  (Methodological)\/}~{\em 39\/}(2), 172--192.

\bibitem[\protect\citeauthoryear{Robberechts, Van~Haaren, and
  Davis}{Robberechts et~al.}{2019}]{robberechts2019will}
Robberechts, P., J.~Van~Haaren, and J.~Davis (2019).
\newblock Who will win it? an in-game win probability model for football.
\newblock arXiv:1906.05029.

\bibitem[\protect\citeauthoryear{Routley and Schulte}{Routley and
  Schulte}{2015}]{routley2015markov}
Routley, K. and O.~Schulte (2015).
\newblock A markov game model for valuing player actions in ice hockey.
\newblock In M.~Meila and T.~Heskes (Eds.), {\em UAI}, pp.\  782--791.

\bibitem[\protect\citeauthoryear{{Stan Development Team}}{{Stan Development
  Team}}{2020}]{cmdstan}
{Stan Development Team} (2020).
\newblock Cmdstan: the command-line interface to stan.
\newblock Version 2.22.1.

\bibitem[\protect\citeauthoryear{Van~Haaren, Hannosset, and Davis}{Van~Haaren
  et~al.}{2016}]{van2016strategy}
Van~Haaren, J., S.~Hannosset, and J.~Davis (2016).
\newblock Strategy discovery in professional soccer match data.
\newblock In {\em Proceedings of the KDD-16 Workshop on Large-Scale Sports
  Analytics}, pp.\  1--4.

\bibitem[\protect\citeauthoryear{Veen and Schoenberg}{Veen and
  Schoenberg}{2008}]{veen2008estimation}
Veen, A. and F.~P. Schoenberg (2008).
\newblock Estimation of space--time branching process models in seismology
  using an em--type algorithm.
\newblock {\em Journal of the American Statistical Association\/}~{\em
  103\/}(482), 614--624.

\bibitem[\protect\citeauthoryear{Vehtari, Gelman, and Gabry}{Vehtari
  et~al.}{2017}]{vehtari2017practical}
Vehtari, A., A.~Gelman, and J.~Gabry (2017).
\newblock Practical bayesian model evaluation using leave-one-out
  cross-validation and waic.
\newblock {\em Statistics and computing\/}~{\em 27\/}(5), 1413--1432.

\bibitem[\protect\citeauthoryear{Vehtari, Gelman, Simpson, Carpenter, and
  Bürkner}{Vehtari et~al.}{2021}]{vehtari2021rank}
Vehtari, A., A.~Gelman, D.~Simpson, B.~Carpenter, and P.-C. Bürkner (2021).
\newblock {Rank-Normalization, Folding, and Localization: An Improved
  $\widehat{R}$ for Assessing Convergence of MCMC (with Discussion)}.
\newblock {\em Bayesian Analysis\/}~{\em 16\/}(2), 667--718.

\bibitem[\protect\citeauthoryear{Wang, Zhu, Hu, Shen, and Yao}{Wang
  et~al.}{2015}]{wang2015discerning}
Wang, Q., H.~Zhu, W.~Hu, Z.~Shen, and Y.~Yao (2015).
\newblock Discerning tactical patterns for professional soccer teams: an
  enhanced topic model with applications.
\newblock In {\em Proceedings of the 21th ACM SIGKDD International Conference
  on Knowledge Discovery and Data Mining}, pp.\  2197--2206.

\end{thebibliography}



\section*{Supplementary material}

# Supporting materials document for Bayesian modelling of marked point processes with applications to event sequences from football

by

Santhosh Narayanan[4], Ioannis Kosmidis[1], and Petros Dellaportas[2,3]

[1]Department of Statistics, University of Warwick, Gibbet Hill Road, Coventry, CV4 7AL, UK
[2]Department of Statistical Science, University College London, Gower St., London, WC1E 6BT, UK
[3]Department of Statistics, Athens University of Economics and Business, 76 Patission Str., Athens, 10434, Greece
[4]The Alan Turing Institute, 96 Euston Road, London, England, NW1 2DB, UK

October 17, 2022

## S1   Derivation of posterior distributions using conjugate priors

### S1.1   Markov chain model for the locations

The probability mass function for the locations specified in expression (9) of the main text, models the locations as a multinomial distribution given the current state (defined by the location and the mark of the last observed event). Similar to the model for the inter-arrival times, the model for the locations is another component of the complete model specification in expression (7) of the main text. We are able to perform inference for this model separately as it does not share any parameters with the other components. Each row of the transition probability matrix $\boldsymbol{\eta}$, corresponding to a single state, is a set of multinomial parameters, one for each location, that add up to 1.

Let $\boldsymbol{y} = \{y_{i \to j}\}$, for $j \in \{1, \ldots, Z\}$, be the observed counts of transitions originating from the state $i$ where $i \in \{1, \ldots, Z\} \times \{1, \ldots, M\}$. Table S1 gives the observed transition counts from the first 5 states in the training data. Out of a total of 90 states, 23 are never observed in the dataset, for example, it is nearly impossible for a Home_Shot event to occur in the defensive third (zone = 1) of the home team.

The likelihood of $\boldsymbol{y}_i$ given the multinomial probabilities $\boldsymbol{\eta}_i$ is

$$p(\boldsymbol{y}_i \mid \boldsymbol{\eta}_i) \propto \prod_{j=1}^{Z} \eta_{i \to j}^{y_{i \to j}},$$

where $\sum_{j=1}^{Z} \eta_{i \to j} = 1$. The conjugate prior for the multinomial distribution is the Dirichlet distribution (see, for example, Gelman et al., 2013, Section 3.4),

$$p(\boldsymbol{\eta}_i \mid \boldsymbol{\nu}_i) \propto \prod_{j=1}^{Z} \eta_{i \to j}^{\nu_{i \to j} - 1},$$

Table S1: Observed transition counts $y_{i \to j}$ from the first 5 states to zones in the training data.

| state $i$ | | next zone $j$ | | |
|---|---|---|---|---|
| zone | mark label | 1 | 2 | 3 |
| 1 | Home_Win | 195 | 38 | 1 |
| 1 | Home_Dribble | 12 | 5 | 0 |
| 1 | Home_Pass_S | 845 | 797 | 51 |
| 1 | Home_Pass_U | 75 | 304 | 160 |
| 1 | Home_Shot | 0 | 0 | 0 |

Table S2: Posterior means of the multinomial transition probabilities $\eta_{i \to j}$ from the first 5 states.

| state $i$ | | next zone $j$ | | |
|---|---|---|---|---|
| zone | mark label | 1 | 2 | 3 |
| 1 | Home_Win | 0.83 | 0.16 | 0.01 |
| 1 | Home_Dribble | 0.65 | 0.30 | 0.05 |
| 1 | Home_Pass_S | 0.50 | 0.47 | 0.03 |
| 1 | Home_Pass_U | 0.14 | 0.56 | 0.30 |
| 1 | Home_Shot | 0.33 | 0.33 | 0.33 |

where $\boldsymbol{\nu}_i > 0$ are the hyperparameters. The posterior distribution of $\boldsymbol{\eta}_i$ is therefore a Dirichlet with parameters $\boldsymbol{\nu}_i + \boldsymbol{y}_i$. To have a non-informative prior we set the hyperparameters $\boldsymbol{\nu}_i = \nu = 1$ and the resulting posterior means of the parameters $\eta_{i \to j}$ are given in Table S2.

### S1.2 Baseline homogeneous Poisson process model

The likelihood for the homogeneous Poisson model for marked spatio-temporal data as specified in Section 5.6 of the main text is

$$\mathcal{L}^{(P)}(\boldsymbol{q} \mid \boldsymbol{\rho}) = \prod_{m=1}^{M} \prod_{z=1}^{Z} \rho_{mz}^{q_{mz}} \exp\left\{-T\rho_{mz}\right\},$$

where $\rho_{mz}$ is the Poisson rate parameter and $q_{mz}$ is the number of event occurrences for mark $m$ at location $z$ over a total observation time $T$ in the data. Table 5 of the main text gives the observed counts $N_{m,z}$ in the training data. The conjugate prior for the Poisson process likelihood is a Gamma distribution

$$p(\boldsymbol{\rho} \mid \kappa, \tau) \propto \prod_{m=1}^{M} \prod_{z=1}^{Z} \rho_{m,z}^{\kappa-1} \exp\left(-\tau \, \rho_{m,z}\right),$$

where $\kappa > 0$ and $\tau > 0$ are the hyperparameters for the shape and rate of the Gamma distribution respectively. Therefore, the posterior distribution of $\boldsymbol{r}$ is a Gamma distribution

$$\kappa' = \kappa + N_{m,z} \qquad \tau' = \tau + T\,,$$

where $\kappa'$ and $\tau'$ are the updated hyperparameters. We set the values, $\kappa = 1$ and $\tau = 0$ that correspond to a non-informative prior.

The resulting posterior means of the Poisson rates $\rho_{m,z}$, for the first 5 marks in each zone, are given in Table S3. We use the rgamma function from the R package stats, which implements the method proposed by Ahrens and Dieter (1982), for simulating from a Gamma distribution.

Table S3: Posterior means of the homogeneous Poisson rates $\rho_{m,z}$, for the first 5 marks.

| mark | | zone | | |
|---|---|---|---|---|
| m | label | 1 | 2 | 3 |
| 1 | Home_Win | 0.0035 | 0.0038 | 0.0006 |
| 2 | Home_Dribble | 0.0003 | 0.0014 | 0.0014 |
| 3 | Home_Pass_S | 0.0251 | 0.0683 | 0.0244 |
| 4 | Home_Pass_U | 0.0080 | 0.0122 | 0.0107 |
| 5 | Home_Shot | 0.0000 | 0.0000 | 0.0043 |

Table S4: Transition counts $c_{i \to j}$ from the first 5 states to the first 5 marks in the training data. We abbreviate the prefix Home to H in the mark labels.

| state $i$ | | label of next mark $j$ | | | | |
|---|---|---|---|---|---|---|
| mark label | zone | H_Win | H_Dribble | H_Pass_S | H_Pass_U | H_Shot |
| H_Win | 1 | 0 | 0 | 80 | 25 | 0 |
| H_Win | 2 | 0 | 8 | 138 | 18 | 0 |
| H_Win | 3 | 0 | 4 | 28 | 11 | 4 |
| H_Dribble | 1 | 1 | 1 | 8 | 3 | 0 |
| H_Dribble | 2 | 0 | 5 | 39 | 11 | 0 |

### S1.3  Baseline Markov chain model for the marks

The probability mass function for the marks specified in expression (15) of the main text, models the marks as a multinomial distribution given the current state (defined by the current location and the mark of the last observed event). Each row of the transition probability matrix $\boldsymbol{\theta}$, corresponding to a single state, is a set of multinomial parameters, one for each mark, that add up to 1.

Similar to the model for locations in Section S1.1, let $\boldsymbol{c} = \{c_{i \to j}\}$, for $j \in \{1, \ldots, M\}$, be the count of observations of the transitions from the state $i$ where $i \in \{1, \ldots, M\} \times \{1, \ldots, Z\}$. Table S4 gives the observed counts of transitions from the first 5 states in the training data.

The likelihood of $\boldsymbol{c}$ given the multinomial parameters $\boldsymbol{\theta}$ is

$$p(\boldsymbol{c}_i \mid \boldsymbol{\theta}_i) \propto \prod_{j=1}^{M} \theta_{i \to j}^{c_{i \to j}},$$

where the sum of the probabilities, $\sum_{j=1}^{M} \theta_{i \to j} = 1$. The conjugate prior for the multinomial distribution is the Dirichlet distribution,

$$p(\boldsymbol{\theta}_i \mid \mathbf{u}_i) \propto \prod_{j=1}^{M} \theta_{i \to j}^{u_{i \to j} - 1},$$

where $\mathbf{u}_i > 0$ are the hyperparameters. The posterior distribution of $\boldsymbol{\theta}_i$ is therefore a Dirichlet with parameters $\mathbf{u}_i + \boldsymbol{c}_i$. We set $\mathbf{u}_i$ to 1 and the resulting posterior means of the parameters $\theta_{i \to j}$ corresponding to the first 5 states are given in Table S5.

## S2  Dealing with model complexity

The conditional mark distribution in the M$\beta$ and M$\beta$A models involves a large number of parameters. There are $M^2 Z$ decay rate parameters and $M(M-1)Z$ baseline conversion rate

Table S5: Posterior means of the multinomial parameters $\theta_{i \to j}$ corresponding to the first 5 states. We abbreviate the prefix Home to H in the mark labels.

| state $i$ | | label of next mark $j$ | | | | |
|---|---|---|---|---|---|---|
| mark label | zone | H_Win | H_Dribble | H_Pass_S | H_Pass_U | H_Shot |
| H_Win | 1 | 0.01 | 0.01 | 0.74 | 0.24 | 0.01 |
| H_Win | 2 | 0.01 | 0.05 | 0.82 | 0.11 | 0.01 |
| H_Win | 3 | 0.02 | 0.10 | 0.56 | 0.23 | 0.10 |
| H_Dribble | 1 | 0.11 | 0.11 | 0.50 | 0.22 | 0.06 |
| H_Dribble | 2 | 0.02 | 0.10 | 0.67 | 0.20 | 0.02 |

parameters which makes posterior sampling a computationally challenging task. We have developed a screening procedure that operates on the data involved in the likelihood and eliminates parameters prior to posterior sampling.

In the M$\beta$ model, the decay rate parameters $\boldsymbol{\beta}$ and conversion rate parameters $\boldsymbol{\gamma}$ capture the duration and magnitude of the excitation effects between all pairs of event types. However, it is reasonable to assume that the matrices $\boldsymbol{\beta}$ and $\boldsymbol{\gamma}$ are sparse, because the excitation effects between all event pairs are not equally significant. To be precise, we expect most elements of the $\boldsymbol{\beta}$ matrix to be infinite, meaning the corresponding excitations decay almost instantaneously. For the $\boldsymbol{\gamma}$ matrix, we expect most its values to be zero, meaning the corresponding event conversions have probability zero. For example, a successful Pass event by one team cannot significantly excite a Pass event for the opposite team, as this would make the commonplace occurrence of a string of passes by a single team very unlikely. If we are able to identify the most significant pairs of event interactions, we can thereby limit the number of elements within the matrices $\boldsymbol{\beta}$ and $\boldsymbol{\gamma}$ that we need to estimate.

## S2.1 Association rule learning

Association rule learning is a method for discovering strong relationships between variables in large databases (see, for example, Agrawal et al., 1993). For example, the association rule Bread $\Rightarrow$ Butter identified from a supermarket sales database would indicate that if a customer buys bread, they are also likely to buy butter. The objective of association rule learning is to identify rules that are interesting based on some measure of significance.

## S2.2 Definition for event sequences

Inspired by the original definition in Agrawal et al. (1993, Section 2), we define the problem of association rule learning in the context of event sequences as

**Definition S2.1.**

- *Let $A = \{1, \ldots, M\}$ be the set of $M$ distinct event types.*

- *Let $B = \{b_{sn}\}$, where $b_{sn} \in A$ for $s \in \{1, \ldots, S\}$ and $n \in \{1, \ldots, N_s\}$, be the training data consisting of $S$ event sequences with $N_s$ number of observed events in the sequence $s$.*

- *Construct a database of subsequences $D = \{d_1, \ldots, d_C\}$, where $C = \sum_1^S N_s$, such that each event $b$ in $B$ has a corresponding subsequence of length $W + 1$ in $D$, made up of $b$ and the $W$ events preceding $b$.*

- *Each subsequence in $D$ is denoted by $d_i = \{x_{i1}, \ldots, x_{iW}, y_i\}$, where $x_{ij}, y_i \in B$ for $i \in \{1, \ldots, C\}$ and $j \in \{1, \ldots, W\}$. We call $\{x_{i1}, \ldots, x_{iW}\}$ as the transient events of the*

Table S6: Support $P(x \cap y)$ for selected event pairs in the training data, where the rows denote the transient event $x$ and columns are the terminal event $y$.

|             | Home_Win | Home_Dribble | Home_Pass_S | Home_Pass_U |
|-------------|----------|--------------|-------------|-------------|
| Home_Win    | 0.0015   | 0.0027       | 0.0663      | 0.0124      |
| Home_Dribble| 0.0006   | 0.0008       | 0.0158      | 0.0030      |
| Home_Pass_S | 0.0111   | 0.0099       | 0.5925      | 0.0962      |
| Home_Pass_U | 0.0163   | 0.0026       | 0.1036      | 0.0289      |

*subsequence before the terminal event $y_i$. Depending on $W$, the elements of the subsequence corresponding to the initial events of a sequence can be empty, because they have shorter histories.*

- *Given a set of event types $A$ and a database of subsequences $D$, a rule is defined as an implication of the form: $x \Rightarrow y$, where $x, y \in A$. The association rule has the interpretation that the event type $x$ is likely to be a transient event in subsequences terminating with event type $y$.*

In other words, the rule $x \Rightarrow y$, would indicate that the event type $x$ excites the occurrence chance of an event with type $y$.

## S2.3  Measures of significance

To identify interesting association rules, we place constraints on two measures of significance (Brin et al., 1997), namely support and lift.

### S2.3.1  Support

The support of $x$ with respect to a rule $x \Rightarrow y$ and a database $D$ is defined as the proportion of subsequences $d$ in the database which contain $x$ as a transient event,

$$P(x) = \frac{|\{d \in D; x \in \mathsf{trans}(d)\}|}{|D|},$$

where $|\cdot|$ denotes the cardinality of a set and $\mathsf{trans}(d)$ is the set of transient events in the subsequence $d$. Similarly, the support of $y$ with respect to a rule $x \Rightarrow y$ is defined as the proportion of subsequences $d$ which terminate with $y$,

$$P(y) = \frac{|\{d \in D; y \in \mathsf{term}(d)\}|}{|D|},$$

where $\mathsf{term}(d)$ is the terminal event in the subsequence $d$.

The support of a rule $x \Rightarrow y$ is defined as, the proportion of subsequences $d$ which contain $x$ as a transient event and terminate in $y$,

$$P(x \cap y) = \frac{|\{d \in D; x \in \mathsf{trans}(d); y \in \mathsf{term}(d)\}|}{|D|}.$$

Table S6 gives the support $P(x \cap y)$ for selected event pairs in the training data.

Table S7: lift($x \Rightarrow y$) for selected event pairs in the training data, where the rows denote the transient event $x$ and columns are the terminal event $y$.

|              | Home_Win | Home_Dribble | Home_Pass_S | Home_Pass_U |
|--------------|----------|--------------|-------------|-------------|
| Home_Win     | 0.4141   | 2.2669       | 0.9793      | 0.9766      |
| Home_Dribble | 0.7176   | 2.7990       | 0.9588      | 0.9698      |
| Home_Pass_S  | 0.3845   | 1.0141       | 1.0879      | 0.9450      |
| Home_Pass_U  | 2.3031   | 1.0860       | 0.7782      | 1.1609      |

### S2.3.2 Lift

The lift of a rule $x \Rightarrow y$ is defined as

$$\text{lift}(x \Rightarrow y) = \frac{P(x \cap y)}{P(x) \cdot P(y)} \,.$$

If the lift of a rule equals 1, it would indicate that the occurrence of $y$ is independent of that of $x$. If the rule has lift $> 1$, then the event $x$ excites the occurrence chance of $y$ and lift $< 1$ indicates $x$ inhibits the occurrence of $y$. Table S7 gives the lift($x \Rightarrow y$) for selected event pairs in the training data.

We implement the following steps to place constraints on the lift and support measures and identify significant dependence between pairs of events.

- Create a database of subsequences as defined in Definition S2.1, for $W = 5$ and $W = 10$, where $W$ is the number of transient events in each subsequence.

- For each $W$, calculate lift for all event pairs and retain only those pairs that have lift $> 1$.

- Set a threshold on the support $P(x \cap y) > \epsilon$, such that when $\epsilon = \epsilon_1$ exactly $N = 50$ event pairs remain, and when $\epsilon = \epsilon_2$, $N = 100$ event pairs remain.

In this way, we select the specific elements of the matrices $\boldsymbol{\beta}$ and $\boldsymbol{\gamma}$, corresponding to the identified significant event pairs, for parameter estimation. The elements of the matrices corresponding to the discarded event pairs are fixed, to the value $10^6$ in the case of the decay rates $\boldsymbol{\beta}$, and $10^{-6}$ for the conversion rates $\boldsymbol{\gamma}$. A large value for the decay rate causes the excitation to die out almost instantaneously, and a very small value for the conversion rate makes the event conversion extremely unlikely. The results of evaluating four separate models, that are fitted based on the specific choices of the tuning parameters given above for the length of subsequence window $W$ and the number of identified event pairs $N$, are discussed in Section 6.2 of the main text.

# References

Agrawal, R., T. Imieliundefinedski, and A. Swami (1993). Mining association rules between sets of items in large databases. *SIGMOD Rec. 22*(2), 207–216.

Ahrens, J. H. and U. Dieter (1982). Generating gamma variates by a modified rejection technique. *Communications of the ACM 25*(1), 47–54.

Brin, S., R. Motwani, J. D. Ullman, and S. Tsur (1997). Dynamic itemset counting and implication rules for market basket data. *SIGMOD Rec. 26*(2), 255–264.

Gelman, A., J. Carlin, H. Stern, D. Dunson, A. Vehtari, and D. Rubin (2013). *Bayesian Data Analysis* (3rd ed.). Boca Raton, Florida: CRC Press.



\end{document}